\documentclass[traditabstract,bibyear]{aa}
\usepackage{times}
\usepackage{natbib}
\usepackage{footmisc}
\usepackage[draft]{hyperref}

\usepackage{epsfig,graphicx,lscape}
\usepackage{float,longtable,comment}
\usepackage{rotating,amssymb,amsmath}
\bibpunct{(}{)}{;}{a}{}{,}

\newcommand{\kms}{{km\,s}$^{-1}$}
\newcommand{\teff}{$T_\mathrm{eff}$\,}
\newcommand{\logg}{$\log g$\,}

\newcommand{\Msun}{\,$\rm{M}_\odot$}
\newcommand{\Lsun}{\,$\rm{L}_\odot$}

\newcommand{\vsini}{$v_{\rm{e}} \sin i$}
\newcommand{\ve}{$v_{\rm{e}}$}

\newcommand{\vi}{$v_{\rm{i}}$}

\newcommand{\Dvr}{$\Delta v_{\rm{r}}$}
\newcommand{\up}{$\leq 40$}
\newcommand{\pmg}{mas\,yr$^{-1}$}

\LTcapwidth=\textwidth

\DeclareRobustCommand{\rchi}{{\mathpalette\irchi\relax}}
\newcommand{\irchi}[2]{\raisebox{0.35ex}{$#1\chi$}} 

\def\5{\footnotesize V\normalsize}
\def\4{\footnotesize IV\normalsize}
\def\3{\footnotesize III\normalsize}
\def\2{\footnotesize II\normalsize}
\def\1{\footnotesize I\normalsize}
\def\lam{$\lambda$}

\def\pp{$\phantom{-}$}
\def\o{$\phantom{1}$}

\begin{document}
	
\title{A census of massive stars in NGC\,346\thanks{Based on observations at the European Southern Observatory in programmes 171.D-0237 and 074D.0011}}
	\subtitle{Stellar parameters and rotational velocities}

\author{P. L. Dufton\inst{\ref{i1}} \and C. J. Evans\inst{\ref{i2}} \and I. Hunter\inst{\ref{i1}} \and D. J. Lennon\inst{\ref{i3}} \and F.R.N. Schneider\inst{\ref{i4},\ref{i5}}}
	
\institute{Astrophysics Research Centre, School of Mathematics \& Physics, The
	Queen's University of Belfast, Belfast, BT7 1NN, Northern Ireland, UK\label{i1} 
	\and UK Astronomy Technology Centre, Royal Observatory, Blackford Hill, 
	Edinburgh, EH9 3HJ, UK\label{i2}
    \and Instituto de Astrof\'isica de Canarias, E-38205 La Laguna, Tenerife, Spain\label{i3}
    \and Zentrum f\"{u}r Astronomie der Universit\"{a}t Heidelberg, Astronomisches Rechen-Institut, M\"{o}nchhofstr. 12-14, 69120 Heidelberg, Germany \label{i4}
    \and Heidelberger Institut f\"{u}r Theoretische Studien, Schloss-Wolfsbrunnenweg 35, 69118 Heidelberg, Germany \label{i5}
}
\date{Received; accepted }
	
\abstract{Spectroscopy for 247 stars towards the young cluster
  NGC\,346 in the Small Magellanic Cloud has been combined with that
  for 116 targets from the VLT-FLAMES Survey of Massive Stars.
  Spectral classification yields a sample of 47 O-type and 287 B-type
  spectra, while radial-velocity variations and/or spectral
  multiplicity have been used to identify 45 candidate single-lined
  (SB1) systems, 17 double-lined (SB2) systems, and one triple-lined
  (SB3) system.  Atmospheric parameters (\teff and \logg) and
  projected rotational velocities (\vsini) have been estimated using
  {\sc tlusty} model atmospheres; independent estimates of \vsini\
  were also obtained using a Fourier Transform method. Luminosities
  have been inferred from stellar apparent magnitudes and used in
  conjunction with the \teff and \vsini\ estimates to constrain
  stellar masses and ages using the {\sc bonnsai} package. We find
  that targets towards the inner region of NGC\,346 have higher median
  masses and projected rotational velocities, together with smaller
  median ages than the rest of the sample. There appears to be a
  population of very young targets with ages of less than 2\,Myr,
  which have presumably all formed within the cluster. The more
  massive targets are found to have lower projected rotational
  velocities consistent with previous studies. No significant evidence
  is found for differences with metallicity in the stellar rotational
  velocities of early-type stars, although the targets in the Small
  Magellanic Cloud may rotate faster than those in young Galactic
  clusters.  The rotational velocity distribution for single
  non-supergiant B-type stars is inferred and implies that a
  significant number have low rotational velocity ($\simeq$10\% with
  \ve$<$40\,\kms), together with a peak in the probability
  distribution at \ve$\simeq$300\,\kms. Larger projected rotational velocity estimates have been found for our Be-type sample and imply that most have rotational velocities between 200--450\,\kms}


\titlerunning{Stellar parameters and rotational velocities in NGC\,346}  
	
\maketitle
%
\section{Introduction} \label{s_intro}

Massive stars significantly influence the evolution of their host 
clusters and galaxies via feedback of both energy and chemically-processed 
material. To help reconcile evolutionary predictions of massive stars 
with observations, the effects of stellar rotation have been included in 
theoretical models \citep{heg00b,mey00} and applied to, for example, 
understanding the ratios of red-to-blue supergiants \citep{mae01} 
and the populations of Wolf--Rayet stars \citep{mey05, vin05}.

One of the primary motivations for including rotation in massive-star
models were observations of core-processed material, for example
enhanced nitrogen abundances, on the stellar surface \citep[see,
e.g.][]{wal70, gie92, ven99, bou03, len03, kor02, kor05}. The process
of rotational mixing then naturally explained how material could be
mixed from the core to the surface. Additionally, rapid rotation may
be a prerequisite in producing a gamma-ray burst from a single massive
star via its homogenous evolution \citep{yoo05, woo06}.

A large observational sample of $\sim$500 OB-type stars in the Galaxy
and Magellanic Clouds was obtained by the FLAMES Survey of Massive
Stars \citep[hereafter FSMS,][]{eva05,eva06} to investigate these
topics.  Surface nitrogen abundances estimated for the B-type samples
in the FSMS \citep{hun07,tru07,hun08a, hun08b} implied rotational
mixing might not be the only transport mechanism involved, and the
nature of these stars has been discussed further by, for example,
\citet{bro11b}, \citet{mae14} and \citet{aer14a}. O-type stars with
low projected rotational velocities and abundances that appear
inconsistent with rotational mixing were also identified by \citet{riv12}.

In a second campaign, the VLT-FLAMES Tarantula Survey \citep[hereafter
VFTS]{eva11}, spectroscopy was obtained for $\sim$800 targets in the
30~Doradus region of the Large Magellanic Cloud (LMC). Targets
with enhanced surface nitrogen abundances, which appeared to be
incompatible with current single-star models of rotational mixing,
were again identified in both the O-type \citep{gri17} and B-type
\citep{duf18} populations. Definitive conclusions have been hampered
both by observational and theoretical uncertainties, and by the
possibility of other evolutionary scenarios, such as binarity and
magnetic fields \citep[see][for further details]{gri17,duf18}.
Nevertheless, rotation remains one of the critical parameters in the
evolution of massive stars.

Theoretical models predict that massive stars at higher metallicity
have stronger line-driven winds, thereby losing more mass and angular
momentum over their lifetime compared to those at lower metallicities
\citep{kud87, kud00, vin05, mok06}. However, studies of rotational
velocities of early-type stars from ultraviolet spectroscopy have
found little compelling evidence for stars in the Clouds rotating more
quickly than in the Galaxy \citep{penny04,pg09}. \citet{hun08a}
presented rotational velocities for $\sim$400 massive stars from
optical spectroscopy in both the LMC and the Small Magellanic Cloud
(SMC) and compared these with Galactic field stars to show that the
predicted trend of rotational velocity and metallicity was observable.
However, the analysis was complicated by the nature of the samples.
Although the target stars in the Clouds were in the direction of young
clusters, much of the sample was composed of field stars. Galactic
studies have implied that stars in clusters rotate faster than those
in the field \cite[see, e.g.][]{str05, duf06, hua06, wol07}, with a
similar trend observed in the LMC \citep{kel04,wol08}. This would be
consistent with the stars in the older field populations having spun
down over their lifetimes. Alternatively, as suggested by
\citet{wol07}, the star formation process may be affected by the
properties of the ambient gas, with stars in clusters being born from
more energetic material and having shorter lived magnetically-locked
accretion discs.

Analysis of the VFTS data has yielded estimates of the rotational
velocity distributions and atmospheric parameters for both the O-type
\citep{ram13, ram15, gri17} and B-type \citep{duf12,duf18,gar17}
samples in the 30~Doradus region in the LMC. Here we present complementary
observations of massive stars in the region of the cluster NGC\,346 in
the lower-metallicity environment of the SMC.  Our observations can be
combined with those for early-type stars previously observed
towards this cluster by the FSMS. By using the same observational
settings and tools as in the analysis of the FSMS and VFTS data we can
minimise any systematic effects in the comparison of their fundamental
parameters.

The observational material is presented in Sect.~\ref{s_obs} and its spectral classification discussed in Sect.~\ref{s_class} and Appendix \ref{s_app}. Sect.~\ref{s_analysis} describes the methods used to estimate the atmospheric parameters, projected rotational velocities, luminosity, masses and ages for our targets, which are then discussed in Sect.~\ref{s_discuss}.

\section{Observations}                                             \label{s_obs}

Spectroscopy was obtained using the Fibre Large Array Multi-Element Spectrgraph \citep[FLAMES,][]{pas02} on the Very Large Telescope at the European Southern Observatory (ESO). The FLAMES--Medusa mode was used to feed light from typically 80 stellar targets (plus sky fibres) per fibre configuration to the Giraffe spectrograph. Our primary aim was to obtain a near-complete spectroscopic census of the O-type and early B-type populations of the NGC\,346 region, with targets selected from the ESO Imaging Survey \citep[EIS,][]{mom01}. Hereafter this spectroscopy will be designated the Survey dataset to distinguish it from the FSMS data. 

\begin{figure}
	\begin{center}
		\vspace{-0.25cm}
		\hspace{-0.85cm}
		\includegraphics[angle=270,scale=0.33]{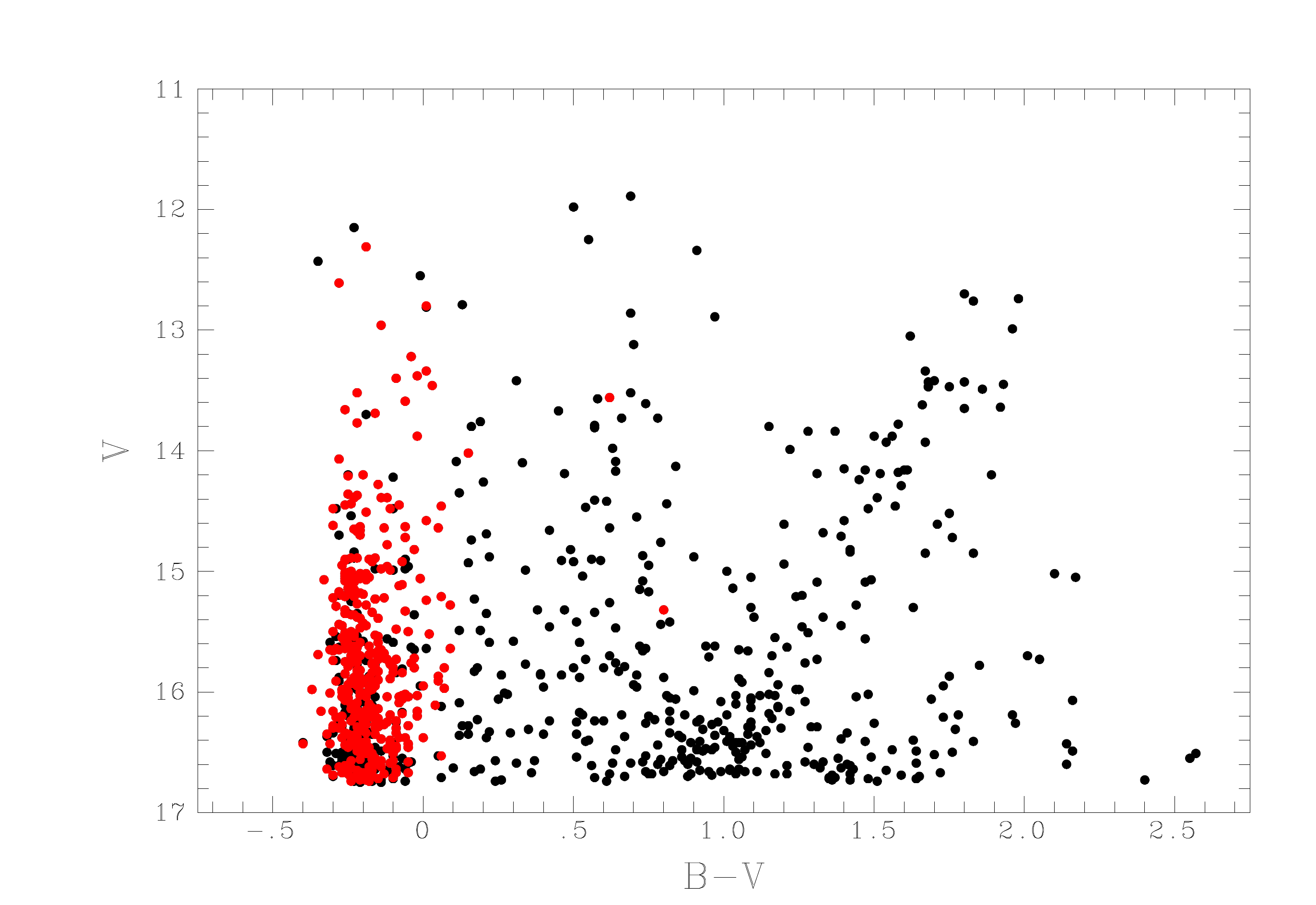}
		\caption{Colour-magitude diagram showing the location of the additional
			FLAMES targets combined with those from the FSMS (363 stars, red
			points) compared to all potential targets ($V$\,$\le$\,16.75) within
			a 10$'$ search radius of the centre of the cluster.}\label{cmd_r10}
	\end{center}
\end{figure}

A faint magnitude cut-off of $V$\,$\le$\,16.75\,mag was employed to ensure
that the signal-to-noise (S/N) ratios of the spectra were sufficient
for quantitative analysis. Adopting a distance modulus to the SMC of
18.9 \citep{har03,hil05}, this corresponds to a latest
(unreddened, main sequence) spectral type of approximately B3~V
\citep[cf.][]{wal72}.  A colour cut of ($B-V$)\,$\le$\,0.1 was also
employed to restrict the sample to OB-type stars, after allowing for a
typical interstellar reddening for NGC\,346 of E($B-V$)\,$\sim$\,0.08
\citep[e.g.][]{hen08}; similar criteria were previously employed in the
selection of the FSMS targets.  A colour-magnitude diagram for our
targets and for the potential targets from the EIS catalogue (within a
10$'$ search radius of the cluster centre) is shown in
Fig.~\ref{cmd_r10}\footnote{Two `red' stars were also observed.  One is
	NGC\,346-003 from the FSMS, observed with the fibre-feed to UVES
	\citep[see][]{eva06}. The other (\#1103) had spurious EIS photometry
	and was later supplemented by archival photometry (see
	Table~\ref{346dat}).}.

Astrometry and optical photometry for our targets are summarised in
Table~\ref{346dat}. Identifications (sorted by $V$-band magnitude)
have been assigned from \#1001 to distinguish them from the FSMS
targets; similar information for the latter is given in Table~4 of
\citet{eva06}. Also included in Table~\ref{346dat} are radial
distances of each target from \#1001 (the brightest object in the
central part of the cluster), following the approach in Table~1 of
\citet{hun08a}. 

\begin{figure*}
	\begin{center}
		\includegraphics[scale=0.62]{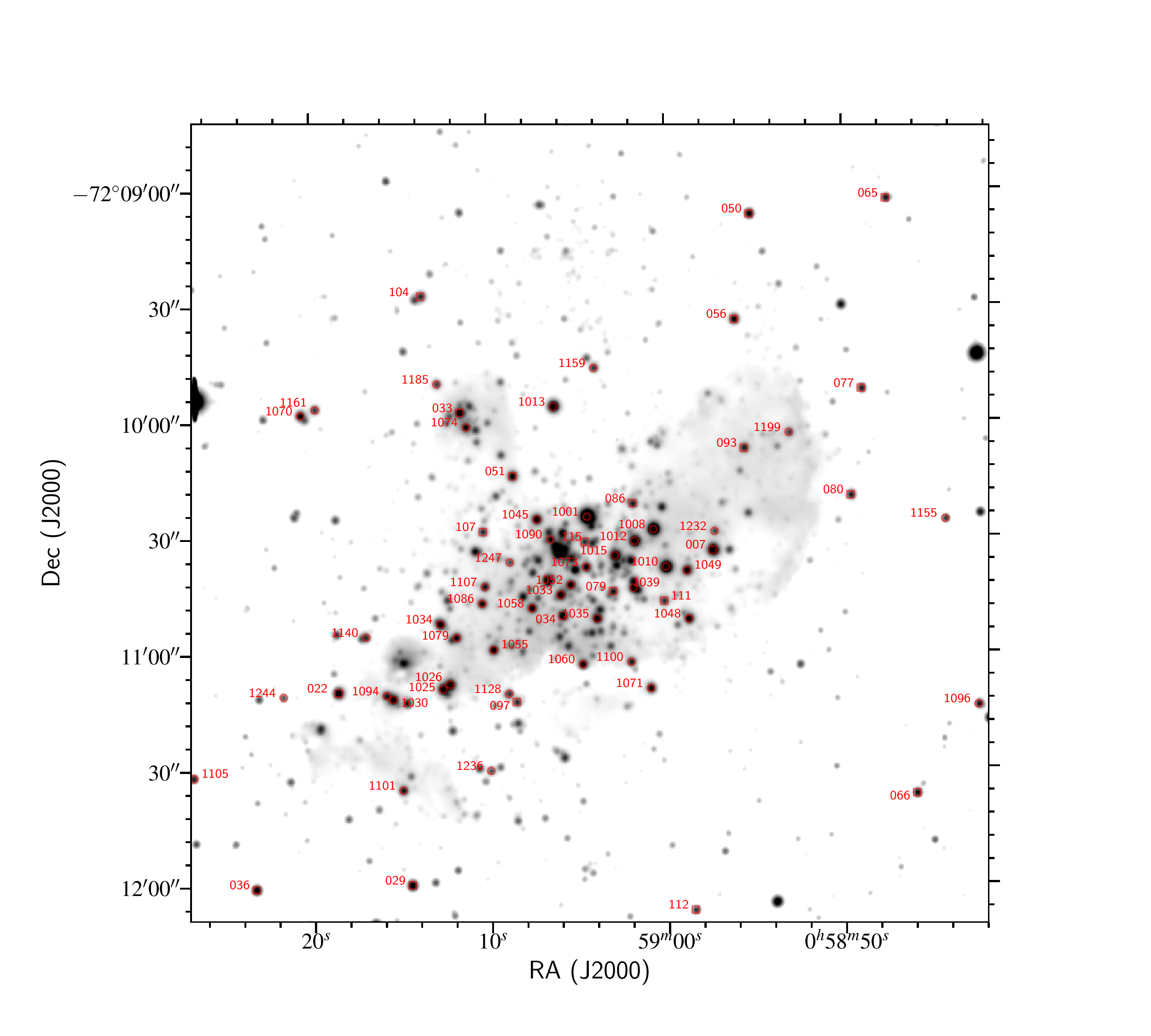}
		\caption{FLAMES targets in the central $\sim$3\farcm5 of NGC\,346. Targets 
			with identifications $>$1000 are from the observations presented here, 
			the remainder are from \citet{eva06}.}\label{fchart1}
	\end{center}
\end{figure*}

The spatial distribution of our targets and those previously observed by the FSMS in the central region of NGC\,346 are shown in Fig.~\ref{fchart1}.  The FLAMES--Medusa fibres project to a diameter of 1\farcs2 on the sky, equivalent to 0.35\,pc in the SMC, and therefore may include contributions from companions unresolved in the ground-based imaging.  From inspection of {\em Hubble Space Telescope (HST)} images of NGC\,346 \citep{sab07}, in all but one instance (\#0111) our targets are by far the brightest source in the fibre aperture. This comparison was only possible for 82 of our 363 sources due to  the limited extent of the {\em HST} imaging. However it includes the densest regions at the centre of the field (Fig.~\ref{fchart1}) where these effects might be expected to be most significant.

The radius of the ionised region of NGC\,346 has been given as
3\farcm5 by \citet{rel02} which, at the distance of the SMC,
corresponds to a physical distance of 61\,pc. Given the magnitude cut-off and colour selection, 125 stars from 167
potential targets were observed out to this radius, a completeness of
75\% which rises to 88\% for $V$\,$<$\,16.0. The principal limitation
on the completeness was the crowding of targets in the core of the
cluster, and the minimum approach distance permitted for the Medusa
fibre-heads. Nearly all of the bright stars in the central region
without FLAMES spectroscopy have previous spectroscopy, leading for
the first time to a comprehensive census of the high-mass
spectroscopic content of NGC\,346.

The observations were taken in service mode between 2004 September 27
and November 28. Three Medusa configurations with near-identical
central positions (Fields `A',`B' and `C') were observed as summarised
in Table~\ref{t_obs}.
Two of the standard Giraffe settings were used:
LR02 (with a wavelength range of 3960 to 4564\,\AA\ at a spectral
resolving power of $R$$\sim$7\,000) and LR03 (4499-5071\,\AA,
$R$$\sim$8\,500). These gave full coverage of the blue optical region,
with an overlap from \lam\lam4500-4565\,\AA. At least four exposures
at each grating setting were taken for each field. In some cases the
requirements on the observing conditions were not satisfied and
observing blocks were repeated. As a consequence, six exposures were
obtained at both settings for Field B, and eight for Field C. Although
the time-sampling is not as extensive as that obtained in the FSMS,
the observations offer some leverage on the detection of massive
binaries.

As for the FSMS data\footnote{For the FSMS targets the normalised spectra were taken directly from the data archive available at: https://star.pst.qub.ac.uk/$\sim$sjs/flames/}, the Giraffe Base-line Reduction Software
\citep[girBLDRS;][]{ble03} was used for bias subtraction, flat-field
correction, fibre extraction and wavelength calibration. The sky
subtraction and correction to the heliocentric velocity frame was then
undertaken using the {\sc starlink} software {\sc dipso} \citep{cur13}.

The multiple spectra of each target were first compared to search for
evidence of radial-velocity variations -- simple division of the
spectra from one epoch into another can reveal `P Cygni'-like features
for binary systems.  The overlap region between the LR02 and LR03
settings helped in this regard as it contains useful stellar lines
(e.g. \ion{He}{ii} \lam4542, \ion{Si}{iii} \lam4552) to increase the
time cadence. Stars displaying radial-velocity variations significant
enough to be detected from this simple approach ($\ga$10\,\kms) are 
considered as candidate single-lined binaries (SB1) in the final column of
Table~\ref{346dat}. Similarly, targets clearly displaying multiple
components are classified as double-lined binaries (SB2), and in one
instance as a triple-lined binary (SB3).
 
For the apparently single stars, all usable spectra were combined either by simple addition or by using a median $\sigma$-clipping algorithm. The final spectra were normally indistinguishable apart from regions affected by cosmic rays where the
latter method was superior. The full wavelength range for spectrum at each grating setting could usually be normalised using a single, low-order polynomial. However, for some features (e.g. the Balmer series), the combined spectra around individual (or groups of) lines were separately normalised.

\citet{gar17} have discussed the reduction of LR02 and LR03 FLAMES spectroscopy for targets which show significant radial-velocity variations. They found that even for narrow-lined stars (with \vsini$\leq$40\kms), simply combining exposures without velocity shifts led to no significant spectral degradation provided the range of radial velocities, \Dvr, was less than 30\,\kms. For targets with larger projected rotational velocities, the maximum value of \Dvr\ increased, e.g.\ for \vsini$\sim$120\,\kms, it corresponded to \Dvr$\leq$100\,\kms.  Hence for the SB1 candidates, we have initially combined the exposures {\em without} any velocity shifts and estimated the \vsini\ as discussed in Sect.\ \ref{s_vsini_method}. We have  then cross-correlated exposures from individual epochs (see Table \ref{t_obs}) to estimate \Dvr\ for each wavelength setting. The \vsini\ and \Dvr\  estimates for each wavelength setting in each target were then compared to the simulations of \citet{gar17}. They were generally consistent with no significant spectral degradation, the only exceptions being: \#1014 (LR02 and LR03 settings), \#1023 (LR03),  \#1101 (LR02),  \#1182 (LR02),  \#1192 (LR03),  \#1196 (LR02),  \#1209 (LR03),  \#1241 (LR03),  \#1246 (LR03). The spectroscopy for these cases has been re-reduced with the individual exposures shifted in velocity space using the results from the cross-correlations. The projected rotational velocities were re-estimated, with the changes in the estimates always being less than 5\%. For the ten additional Survey targets identified as SB2, the exposures for each epoch were combined separately.

\begin{table}
\begin{center}
\caption{Observing log of the FLAMES NGC\,346 Survey observations.}\label{t_obs}
{\small
\begin{tabular}{lcccc}
			\hline\hline
			Date             & Field & Setting & \lam$_c$ & Exposure time \\
			&         &             & [\AA]       & [sec] \\
			\hline 
			2004-09-27 & A & LR02    & 4272   & 2$\times$1725 \\ 
			2004-09-28 & A & LR02    & 4272   & 2$\times$1725 \\ 
			2004-10-01 & A & LR03    & 4797   & 2$\times$1725 \\
			2004-10-04 & A & LR03    & 4797   & 2$\times$1725 \\
			\\
			2004-10-14 & B & LR02    & 4272   & 1$\times$1725, 1$\times$1234 \\ 
			2004-10-17 & B & LR02    & 4272   & 2$\times$1725 \\ 
			2004-10-18 & B & LR03    & 4797   & 4$\times$1725 \\
			2004-11-25 & B & LR02    & 4272   & 2$\times$1725 \\
			2004-11-25 & B & LR03    & 4797   & 2$\times$1725 \\
			\\
			2004-10-18 & C & LR02    & 4272   & 4$\times$1725 \\ 
			2004-10-18 & C & LR03    & 4797   & 4$\times$1725 \\
			2004-11-26 & C & LR02    & 4272   & 4$\times$1725 \\ 
			2004-11-26 & C & LR03    & 4797   & 2$\times$1725 \\ 
			2004-11-28 & C & LR03    & 4797   & 2$\times$1725 \\ 
			\hline
\end{tabular}
}
\end{center}
\end{table}

\section{Spectral classification}\label{s_class}

All the Survey spectroscopy has been classified with reference to criteria 
developed for massive stars in the low metallicity environment of the SMC
\citep{len97,wal00,eva03,eva04}.  Spectral classifications for all 247
targets are given in Table~\ref{346dat}\footnote{These were previously
	included in the compendium of SMC data by \citet{bon10} as
	`Hunter et al. in prep.'}, together with previous published
classifications.  Cross-matches with past identifications were
generally performed by visual comparison of published finding charts
and the EIS images. In the case of the 2dF spectroscopy from
\citet{eva04}, stars were matched using their astrometric data and a
conservative search radius of 1$''$ (to avoid specious matches).
Previous classifications were principally from \citet[][19
stars]{mas89}, and \citet[][18 stars]{eva04} and are included in the
final column of Table~\ref{346dat} in italics. Other classifications
are included as footnotes to the table. Detailed comments on
five targets of note are provided in Appendix \ref{s_app}.

In general there was good agreement with previous classifications, although a few stars (e.g. \#1048 and
\#1058) received later types than before, probably owing to the
improved spectral resolution of the FLAMES data. While good nebular
subtraction is difficult to achieve in fibre spectroscopy, \ion{He}{i}
nebular emission was typically resolved in the cores, highlighting the
potential for infilling if degraded to lower resolution. We were also
able to refine some of the previously uncertain classifications from
\citet{eva04}.

The Survey data thus provide the first classifications for over 200
massive stars in this important region in the SMC, including ten new
O-type stars (six of which are within 1\farcm2 of the centre).  When
combined with the FSMS spectroscopy from \citet{eva06}, this
represents a detailed spectral census of the NGC\,346 cluster and
its environs. A summary of the spectral content of the two
observational samples is given in Table~\ref{types}.

Twelve targets have been classified as O-type binaries (9 SB1, 2
SB2, 1 SB3) and fifty one as B-type binaries (36 SB1, 15 SB2). This
translates into observed binary percentages (plus standard deviations
assuming binomial statistics) of 26$\pm$6\% (O-type) and 18$\pm$2\%
(B-type). For the VFTS survey of the Tarantula Nebula in the LMC,
\textit{observed} binary percentages of 35$\pm$3\%
\citep[O-type][]{san13} and 25$\pm$2\% \citep[B-type][]{dun15} were
found. However, as discussed by those authors, the actual binary
percentages will be larger due to some binaries not having been
identified; from Monte-Carlo simulations they inferred \textit{actual}
binary percentages of more than 50\% for both O- and B-type systems.
Given the limited time cadence of our spectroscopy, our
\textit{observed} binary fractions appear consistent with those found
from the VFTS.

Known omissions from our spectroscopy (due to crowding of the Medusa
fibres) in the central part of the cluster include: MPG\,395, 451,
468, and 487, with classifications from \citet{mas89} of B0~V, B0~V,
O9~V, and O6.5~V, respectively; MPG\,470, classified as O8-9 III:nw
\citep{wal86a}; the components of N66A, which include N66A-1
\citep[O8~V,][]{hey10}; and MPG\,375 ($V$\,$=$\,15.47), for which the
spectral type is unknown.  With our focus on OB-type stars, we
have also omitted HD\,5980, the bright source on the eastern edge of
Fig.\ \ref{fchart1}, approx. 1\farcm75 from the core of NGC\,346.
HD\,5980 is a well-studied WR/LBV binary system, with some recent
observations suggesting it it is potentially a quadruple system
\citep{koe14}.

\begin{table}
	\begin{center}
		\caption{Spectral content of the NGC\,346 region from the FSMS \citep{eva06}
			and the Survey data. All of the Be-type spectra are of
			early B-type, except for two classified as B5 in the Survey dataset. \#1024 has been classified as a Be-type star but see Appendix \ref{s_app} for a detailed discussion of its spectrum.}\label{types}
		\begin{tabular}{lcccccc}
			\hline\hline
			Source              & Total     & O        & Early-B   & Be  & Late-B & AFG \\
			&           &          & [B0-3]    &          & [B5-9] &     \\
			\hline
			%
			%
			FSMS  & 116       & 19       & \o59      & 25       & 2      & 11  \\	
			%
			%
			Survey          & 247       & 28       & 149       & 45       & 8      & 17 \\  
			Total               & 363       & 47       & 208       & 70       & 10     & 28 \\
			\hline
		\end{tabular}
	\end{center}
\end{table}

\section{Analysis}                                            \label{s_analysis}

\subsection{Atmospheric parameters}  \label{s_atm}

We have employed model-atmosphere grids calculated with the {\sc tlusty} 
and {\sc synspec} codes \citep{hub88, hub95, hub98, lan07}.
They cover a range of effective temperature, 12\,000
$\leq$\,\teff$\,\leq$\,35\,000\,K in steps of typically 1\,500\,K.
Logarithmic gravities (in cm\,s$^{-2}$) range from 4.5 dex down to the
Eddington limit in steps of 0.25 dex, and microturbulences are from
0-30 \kms\ in steps of 5 \kms. Grids have been calculated for a range
of metallicities with that for an SMC metallicity used here. As
discussed by \citet{rya03} and \citet{duf05}, equivalent widths and
line profiles interpolated within these grids are in good agreement
with those calculated explicitly at the relevant atmospheric
parameters. Full details of the grids can be found in \citet{duf05}.

The analysis followed similar methods to those used by \cite{hun08a}
when analysing the FSMS data and is only briefly summarised here.
Where two (or more) ionisation stages of silicon were observed, the
effective temperatures were constrained by requiring that each
ionisation stage yielded the same estimated silicon abundance. For the
B0 to B0.5 spectral types, the \ion{He}{ii} \lam4542 and \lam4686
lines were also observable and yielded estimates that were normally in good
agreement with those from the silicon ionisation equilibrium.  For
spectral types later than B3, the strength of the \ion{He}{i} spectrum
becomes temperature sensitive and the well observed \lam4026 line was
used.

Effective temperature estimates for the remaining stars have been
taken from the effective temperature--spectral type calibration for
the SMC of \cite{tru07}. We find that this calibration yields
estimates that are in satisfactory agreement with those from the silicon and
helium lines, with a mean difference of 460\,$\pm$\,1220\,K. Our
sample contained targets with B2.5~V and B3~V types, which lie outside
this calibration and for which we have no independent estimates from
the silicon and helium lines. The effective temperature scales from
\citet{tru07} for the Galaxy and the SMC differ by typically
2\,500-3\,500\,K and between the LMC and SMC by 500\,K. We have
therefore used these differences to estimate effective temperatures
for the B2.5~V and B3~V types in the SMC. Finally, the effective
temperature for two B2 II objects was taken as the mean of those for
B2 I and B2 III. The adopted effective temperature calibration versus
spectral type is summarised in Table~\ref{t_tempscale}.

In cases where we were unable to assign unique spectral types, we have
assigned effective temperatures appropriate to the mid-points of the
range. For spectra without luminosity classifications we have assumed
that they are near main-sequence (class V) objects. As discussed in
Section~\ref{s_lum}, the estimated luminosities are generally
consistent with such a classification.

Surface gravities were determined by comparing rotationally-broadened
theoretical profiles of the H$\delta$ and H$\gamma$ lines to those
observed. The surface gravity and effective temperature estimates are
correlated and an iterative method was adopted to simultaneously
determine these parameters. It should be noted that, as discussed by
\citet{dun11}, surface gravity estimates for the Be-type stars may be
too small due to continuum contamination from a circumstellar disc. 

The analyses presented here have implicitly assumed that the targets are single, although a signifiant number of the apparently single targets may be the primary in a multiple system. Additionally we have estimated atmospheric parameters for 36 SB1 and 11 SB2 B-type systems. \citet{gar17} considered the consequences of an unseen secondary in their analysis of VFTS B-type binaries. Making the extreme assumption that the secondary had a featureless spectrum, they found that the estimated effective temperatures (from the silicon ionization equilibrium) and the gravities (from fitting hydrogen line profiles) were too small by typically 500--1\,000\,K and 0.1--0.3\,dex respectively. In reality, the situation will be more complicated as any secondary making a significant flux contribution is also likely to be a near main sequence B-type star but of slightly later spectral type than the primary. If its absorption features were incorporated into those of the primary (as is very likely for the broad hydrogen lines), the consequences could be different to those modelled by \citet{gar17}. For example, as the hydrogen line spectrum in B-type stars strengthens as one moves to later spectral types, the gravity estimates could become {\em too large}. In summary, the atmospheric parameters for all the targets may contain additional uncertainties due to the presence of secondaries. However the discussion above implies that these are unlikely to be significantly larger than the stochastic uncertainties discussed below.

Twenty eight O-type stars were also observed, which are generally
hotter than the models in our {\sc tlusty} grid. Four have been
analysed previously by \citet{bou13}
for \#1008, \#1012, and \#1071\footnote{Estimates for \#1008 and \#1012
have also been presented by \citet{bou03}, \citet{hea06}, and
\citet{mas09}.}, and \citet{hea06} for \#1019 and their estimates are listed in Table~\ref{t_346results}. For the remainder
we adopt temperatures using the calibration for O-type dwarfs from the
analysis of $\sim$30 O-type stars in the SMC by \citet[][which
included results for 21 of the stars observed in NGC\,346 by the FSMS
and is summarised in Table~\ref{t_tempscale}]{mok06}. Seven of the targets are designated as SB1 and the atmospheric parameters for these (and indeed) other O-type targets may be affected by the presence of unseen secondaries as discussed for the B-type sample. Given the complications of modelling the winds of O-type stars we do not attempt
to derive surface gravities or wind parameters for these stars here.

The atmospheric parameters of our sample of stars are given in the
Table~\ref{t_346results}. For completeness, we also list the
parameters from \citet{hun08a} and \citet{mok06} for the FSMS sample \footnote{\citet{hun08a} did not provide atmospheric parameters
for \#0111 although the \ion{He}{ii} spectrum is clearly present in
the FSMS spectroscopy. We have therefore analysed this spectroscopy
to obtain atmospheric parameters, although we note that the {\em HST}
imaging reveals this to be a visual composite of two nearly equal magnitude
sources.}; when no atmospheric parameters were provided, we have
followed the same spectral type methodology as for the Survey spectroscopy. Effective temperatures (and gravities) could not be estimated for 15 targets. Six were SB2 systems where the spectral type of the primary was uncertain, seven lay beyond our spectral type calibration (five B9 II and two B5e targets), and \#1001 and \#1024 had peculiar spectra (as discussed in Sects.\ \ref{1001} and \ref{1024}, respectively). Additionally, no gravity estimate is given for \#1038 (B0:e) as no convincing fit could be obtained for its Balmer line profiles.

Comparison of the effective temperatures derived from the silicon and
helium lines for our sample and for B-type samples obtained with the
same instrumentation in the LMC \citep{gar17, duf18} imply that these
estimates will have a stochastic uncertainty of typically
$\pm$\,1\,000\,K. The effective temperature estimates deduced from the
spectral type calibration will be more uncertain. The comparison with
those estimated from the silicon and helium lines discussed above and
the scatter in effective temperatures estimates for stars of the same
spectral type implies that where the spectral type is well defined a
conservative stochastic uncertainty of $\pm$1\,500\,K is appropriate.
For those stars with a range of spectral types or no luminosity
classification the uncertainties may be larger. For the surface
gravity estimates, the values from the H$\delta$\ and H$\gamma$\
lines generally agreed to within 0.1\,dex. Taking into account the
uncertainties in the effective temperature estimates, an error of
0.2\,dex in the surface gravity would appear to be appropriate.

\begin{table}
	\begin{center}
          \caption[]{Adopted effective temperature--spectral type
            calibration.}\label{t_tempscale}
		\begin{tabular}{lccc}
			\hline \hline
			Spectral type  & \multicolumn{3}{c}{Effective temperature} \\
			& I                 &	    III      & V	      \\
			\hline \\
			O5             &    --             &	--	         & 45200 (M)\,\,\, \\
			O6             &    --             &	--           & 42970 (M)\,\,\, \\
			O7             &    --             &	--           & 40730 (M)\,\,\, \\
			O8             &    --             &	--           & 38500 (M)\,\,\, \\
			O9             &    --             &	--           & 36265 (M)\,\,\, \\
			O9.5           &    --             &	--           & 35150 (M)\,\,\, \\
			\\
			B0             & 27200 (T)\,\,\,   & --              & 32000 (T)\,\,\, \\
			B0.2           & 25750 (T)\,\,\,   & --              & 30800 (T)\,\,\, \\
			B0.5           & 24300 (T)\,\,\,   & --              & 29650 (T)\,\,\, \\
			B0.7           & 22850 (T)\,\,\,   & 25300 (T)\,\,\, & 28450 (T)\,\,\, \\
			B1             & 22350 (T)\,\,\,   & 23950 (T)\,\,\, & 27300 (T)\,\,\, \\
			B1.5           & 20650 (T)\,\,\,   & 22550 (T)\,\,\, & 26100 (T)\,\,\, \\
			B2             & 18950 (T)\,\,\,   & 21200 (T)\,\,\, & 24950 (T)\,\,\, \\
			B2.5           & 17200 (T)\,\,\.   & 19850 (T)\,\,\, & 23900 (E)\,\,\, \\
			B3             & 15500 (T)\,\,\,   & 18450 (T)\,\,\, & 21500 (E)\,\,\, \\
			B5             & 13800 (T)\,\,\,   & --              &  --        \\
			\\ \hline
\end{tabular}
\tablefoot{The calibration for O-type dwarfs was assumed to be linear
  and was derived from fitting the results from \citet[][M]{mok06}.
  The B-type values were primarily from \citet[][T]{tru07} and
  extended by scaling from Galactic and LMC estimates from
  \citeauthor{tru07} (E).}
	\end{center}
\end{table}

\subsection{Projected rotational velocities}              \label{s_vsini_method}

Projected rotational velocities have been estimated using two
independent methodologies, viz. profile fitting (PF) and Fourier
Transform (FT). For the former, the {\sc tlusty} theoretical model at
the closest grid point to the parameters given in
Table~\ref{t_346results} was adopted. An absorption line profile was
then scaled to have the same strength as that observed and the
instrumental broadening was included by convolving with a Gaussian
profile. The resulting profile was then rotationally broadened assuming a linear limb darkening law with $\epsilon$=0.6 \citep{gra05} until
the best fit (by a $\rchi^2$\ minimisation) with the observed profile
was found. This method was used previously by \citet{hun08a} where
further details can be found.

Metal absorption lines, which are less affected by intrinsic
broadening (than for example the hydrogen and diffuse helium lines)
provide the most reliable estimates. For our B-type stars we have used
either the \ion{Mg}{ii} \lam4481 or \ion{Si}{iii} \lam4552 lines,
whichever was stronger. For stars with significant rotational
velocities the metal lines were not well defined, and we used the
\ion{He}{i} \lam4026 line. The rotational broadening of our targets
was generally larger than the intrinsic broadening, so the choice of
theoretical profile was not critical.

For the O-type stars we have adopted theoretical profiles from the
grid of \citet{lan03}. The mean surface gravity of the O-type dwarf
stars from \citet{mok06} was \logg$\sim$4.15\,dex and we therefore
adopted models from the closest effective temperature grid point with
\logg\,$=$\,4.25\,dex.  Our targets have spectral types later than O5
and hence the \ion{He}{i} \lam4026 line was visible and was used to
estimate projected rotational velocity, thereby maintaining
consistency with analysis of the B-type stars. The PF estimates are listed
in Table~\ref{t_346results} with those from the original FSMS data
being taken directly from \citet{hun09a}\footnote{For nine FSMS
  targets, where no values were tabulated, PF estimates have been
  obtained as discussed above.}.

Independent estimates were obtained using the FT methodology
\citep{car33, sim07}. This has been widely used for early-type stars
\citep[see, for example][]{dufsmc06, lef07, mar07, sim10, sim17,
	fra10, duf12, sim14} and relies on the convolution theorem
\citep{gra05}, viz. that the Fourier transform of convolved functions
is proportional to the product of their individual Fourier Transforms.
It then identifies the first minimum in the Fourier transform for a
spectral line, which is assumed to be the first zero in the Fourier
transform of the rotational broadening profile with the other
broadening mechanisms exhibiting either no minima or only minima at
higher frequencies. Further details on the implementation of this
methodology are given by \citet{sim07} and \citet{duf12}. Estimates
were obtained for the targets observed here and those in the FSMS and
are listed in Table~\ref{t_346results}. For the SB2 systems the estimates refer to the primary, which is defined as the star having the strongest absorption spectrum. Values were only measured for SB2 systems where the two spectra were well separated in at least one epoch.

The moderate spectral resolving power of our LR02 and LR03 spectra
corresponds to a velocity resolution of approximately 40\,\kms. This
made the estimation of the projected rotational velocities in the
sharpest lined stars unreliable. For the PF methodology, the
instrumental broadening dominated the observed profiles, while
estimates also became more sensitive to the intrinsic profile adopted.
For the FT methodology, the position of the first minima (at
relatively high frequencies in the Fourier Transform) became difficult
to identify. Hence when projected rotational velocity estimates were
less than 40\,\kms, they have been assigned to a bin with
$0\le$\,\vsini\,$\le40$\ \kms. For consistency the same approach has
been adopted for the previously published PF estimates for the FSMS
data. Sixty one targets fell into this category, with 54 targets
having estimates of $\leq$40\,\kms\ using both methodologies.  One
target had no FT estimate, while for the remaining six objects the
larger (three PF and three FT) estimates ranged from 43-51\,\kms\
implying that they were consistent within the uncertainties discussed below.

For the 47 O-type stars, PF estimates could be obtained for all
targets, apart from \#1010, which has been classified as SB3 with
asymmetric profiles; additionally, no convincing minimum was found for
\#1030 using the FT methodology. Estimates with
\vsini\,$\geq$40\,\kms\ were obtained for 36 targets and yielded a
mean difference (FT-PF) of 4$\pm$16\,\kms\ and a mean ratio (FT/PF) of
1.04$\pm$0.11.

For the larger B-type sample of 288 targets, 275 PF estimates were
available. For those without measurement: five targets were classified
as B9 II and lay beyond the low effective temperature limit of our
grid, seven targets were classified as SB2, while no convincing fit could be found for \#1024 (see Sect.~\ref{1024}). For the FT methodology, 274 estimates
were obtained; again no estimates were obtained for the SB2, B9 II
targets and \#1024, while no convincing minima could be found in the SB2 target, \#0035.  Estimates with \vsini\,$\geq$40\,\kms\ for both
methodologies were obtained for 223 B-type targets and yielded a mean
difference of 1\,$\pm$\,12\,\kms\ and a mean ratio of
1.01\,$\pm$\,0.09.  The statistics for the FSMS (61 targets,
$-$1\,$\pm$\,11\,\kms, 0.99\,$\pm$\,0.07) and Survey (162
targets, 2\,$\pm$\,13\,\kms, 1.02\,$\pm$\,0.09) were similar.

These statistics imply that the estimates from the two methodologies
are in good agreement. The larger errors for the O-type sample may be
due to the intrinsic weakness of the \ion{He}{i} spectra and/or the
larger macroturbulences that have been inferred for these spectral types
\citep{sim14, sim17}. We have adopted conservative stochastic
uncertainties of $\sim$10\% for the larger estimates and $\pm$10\,\kms\
for the estimates with \vsini$<100$\,\kms.

\subsection{Luminosities} \label{s_lum}

Luminosities have been calculated using the methodology described by
\citet{hun07}. A uniform reddening of $E(B-V)$\,$=$\,0.09
\citep{mas95} with a reddening law of $A_{\rm V}$\,=\,2.72$E(B-V)$
\citep{bou85} and a distance modulus of 18.91\,dex \citep{hil05} were
adopted. Bolometric corrections from \citet{vac96} and \citet{bal94}
were used for stars hotter and cooler than 28\,000\,K, respectively.
Luminosity estimates were estimated for all targets apart from the 15
stars without effective temperature estimates (see Sect. \
\ref{s_atm}) and are listed in Table~\ref{t_346results}. The main
sources of uncertainty will arise from the bolometric correction and
the extinction. We estimate that these will typically contribute a
stochastic error of $\pm 0^{\mathrm{m}}$\!\!.5 in the absolute
magnitude corresponding to $\pm$0.2 dex in the luminosity. For
  stars in the innermost part of the cluster the adopted PSF-fitting
  photometry might have been influenced by the diffuse light from
  unresolved fainter stars. This is generally less than 10\% 
  compared to the magnitudes of our sources and less significant than
  the uncertainties already discussed.  Additionally, there may be a
  systematic error due to the adopted distance to the SMC.
  \citet{hil05} estimated an uncertainty in their distance modulus of
  $\sim$0.1, which would translate to a systematic error in the
  luminosity estimates of 0.04\,dex.

As discussed in Sect.\ \ref{s_atm}, for the 66 targets with no luminosity class, 
an effective temperature appropriate to a luminosity class V was adopted. Most of these have estimated luminosities consistent with them being close to the main sequence. However 13 targets have luminosities that are more than 0.3\,dex larger than this main sequence luminosity (identified by a linear fit between luminosity and effective temperature for all targets with a luminosity class V designation). Hence the use of a luminosity class V calibration to estimate the effective temperature may not be appropriate; we have identified these targets in Table \ref{t_346results} and their physical parameters should be treated with caution.

\subsection{Masses and ages} \label{s_mass}

We have used {\sc bonnsai}\footnote{The {\sc bonnsai} web-service is available at: \newline \url{www.astro.uni-bonn.de/stars/bonnsai}.} to estimate the evolutionary masses and ages of the stars in our samples. {\sc bonnsai} uses a Bayesian methodology and the grids of models from \citet{bro11a} to constrain the evolutionary status of a given star, including its age and mass \citep[see][for details]{sch14}. As independent prior functions, we adopted the SMC metallicity grid of models, a \citet{sal55} initial mass function, the initial rotational velocity distribution estimated by \citet{hun08a}, a random orientiation of spin axes, and a uniform age distribution. The estimates of effective temperature, luminosity and FT projected rotational velocity (taken from Table~\ref{t_346results}) were then used to constrain masses. For all targets, the predicted current and initial masses were very similar with differences $<$5\% and in Table \ref{t_346results}, we have therefore only listed the current mass estimates. 

Using the adopted errors on the effective temperature (see Sect.\ \ref{s_atm}), luminosity (Sect.\ \ref{s_lum}) and projected rotational velocities (Sect. \ref{s_vsini_method}), {\sc bonnsai} returned 1$\sigma$-uncertainties for all the quantities that it estimates. In the case of the stellar masses, these were generally 6-8\% and never greater than 10\%. For the ages, the errors were normally 10-15\% (and always less than 20\%) for targets with estimated ages greater than 5\,Myr. For younger targets the absolute error in the age estimates was typically 1-2\,Myr. The corresponding larger fractional uncertainty reflects the position of the targets close to the zero age main sequence (ZAMS) and indeed for the youngest targets the lower error bound was consistent with the target lying on the ZAMS. 

There may be additional uncertainties due to binarity or line-of-sight
composites (unresolved at the distance of the SMC). These could affect the effective temperature estimates as discussed in Sect.\ \ref{s_atm} and also lead to an overestimate of the luminosity for the primary. We have investigated the consequences of this by {\em arbitrarily} decreasing the luminosity estimates for \#0043 (B0 V) and \#0099 (B2 III) by 0.2 dex (corresponding to a secondary flux contribution of approximately 35\%); these apparently single stars were chosen as their effective temperature estimates lie at the upper and lower ranges of our B-type sample. Mass estimates were reduced by $\sim$10\% for both stars with the age estimates decreasing by 10\% and increasing by 15\% respectively. Hence even for apparently single stars care must be taken when interpreting these estimates. 
	
For the SB2 target, \#0013, {\sc bonnsai} mass estimates can be compared with those deduced from the orbital analysis by \citet{ric12}, The {\sc bonnsai} estimate in Table \ref{346dat} is based on an uncertain spectral type (B1:) for the brighter component and a luminosity estimate that has not been corrected for binarity. As such it is larger than that found for the brighter component by \citet[11.9$\pm$0.6\Msun]{ric12}. We can also compare mass estimates based on the stellar parameters given by \citet{ric12} and our estimated system luminosity. For the brighter component, this leads to an effective temperature of 24\,500$\pm$1\,500, $\log$ L/\Lsun\ of 4.67$\pm$0.2\,dex and \vsini\ of 110$\pm$10\,\kms; the values for the secondary component are 34\,500$\pm$3\,000\,K, 4.50$\pm$0.2\,dex and 320$\pm30$\,\kms\  respectively. {\sc bonnsai} returns mass estimates of 12.8$^{+1.9}_{-1.7}$ and 16.0$^{+2.4}_{-2.1}$\,\Msun. compared with 11.9$\pm$0.6 and 19.1$\pm$1.0\,\Msun from \citet{ric12}. Within the error bars, the estimates are in reasonable agreement, and indeed decreasing the adopted luminosity ratio would improve the agreement for both components. This is consistent with the good agreement that \citet{sch14} found in their {\sc bonnsai} analysis of Galactic binaries.

Estimates could not be obtained for 22 targets. Fifteen had no effective temperature estimates (see Sect.\ \ref{s_atm}) and hence no luminosity estimates as their bolometric corrections were unknown. As such they had insufficient constraints to estimate masses or ages. The remaining seven targets failed the posterior predictive check and/or a $\rchi^2$-test. These were all classified as supergiants - four B8Ib/II, and three B1.5-B3 spectral types. For the former it was also not possible to find solutions using the effective temperature, gravity and \vsini\ estimates as constraints, implying that the models of \citet{bro11a} did not cover these stellar parameters. For the three remaining targets, solutions could be found using these constraints but implied logarithmic luminosities (in solar units) of 5.83-5.95\,dex. These are far higher than those found in Sect.\ \ref{s_lum} and may be related to a mass discrepancy discussed by, for example, \citet{her92} and \citet{mar18}. Hence we have not included these mass and age estimates in Table \ref{t_346results}.

In Sect.\ref{s_vsini_method}, projected rotational velocities were estimated for our sample and in Sect.\ref{d_ve}, these are deconvolved to estimate the rotational velocity distribution. This is similar to that estimated by \citet{hun08a} but shows evidence for a double peaked structure. Previously \citet[][Appendix A]{sch17a} found that the different choices of the rotational velocity distribution could lead to significant differences in the mass and age estimates. 

Hence in order to investigate the sensitivity of our estimates to the adopted distribution, we have considered two alternatives -- a flat distribution (with any rotational velocity being equally probable) and that found for B-type stars in 30~Doradus, which shows similar evidence for a double peaked structure (see Fig.\ \ref{f_v_e}). Targets with projected rotational velocities of approximately 100, 200 and 300\,\kms and effective temperatures of approximately 20\,000 and 30\,000\,K were analysed with the results summarized in Table \ref{t_Mt_ve}. In general, different rotational velocity distributions lead to very similar estimates of the stellar mass and luminosity. Indeed in over half the cases, there is no change in the estimates as {\sc bonnsai} selected the same evolutionary models. The maximum ranges in mass and age estimates are only 0.2\,\Msun\ and 0.5\,Myr respectively. As these are both lower than the uncertainties in the estimates discussed above, we conclude that uncertainties due to the choice of rotational velocity distribution are unlikely to be significant at least for targets near the hydrogen burning main sequence.

\begin{table*}
	\begin{center}
		\caption{Estimates of the mass (in units of \Msun) and age (in Myr) for different choices of the rotational velocity distribution. Estimates are provided for the adopted distribution of \citet{hun08a} (H08), a flat distribution (Flat) and the distribution found by \citet{duf12} for B-type stars in 30~Doradus (30~Dor) and are given to one decimal place to aid comparisons. Other stellar parameters are taken from Table \ref{t_346results}.}\label{t_Mt_ve}
		\begin{tabular}{lccrrrrrrr}
		\hline\hline
		Star  & \teff   &  $\log L$ & \vsini &\multicolumn{2}{c}{H08} & \multicolumn{2}{c}{Flat} & \multicolumn{2}{c}{VFTS} \\
			  &         &           &        & ~~~Mass &  Age & ~~~Mass &  Age  &  ~~~Mass &  Age      \\
	    0099  & 18\,000 &   3.79    &   98   &  7.0 & 36.8 &  7.0 & 36.8 &  7.0 & 36.8  \\
	    0070  & 30\,500 &   4.43    &  102   & 13.8 &  7.9 & 13.8 &  7.8 & 13.8 &  7.7  \\
	    \\
	    1088  & 19\,850 &   4.06    &  184   &  8.6 & 24.7 &  8.8 & 24.7 &  8.6 & 24.7  \\
	    1230  & 29\,650 &   4.06    &  198   & 11.8 &  3.9 & 11.8 &  3.4 & 11.8 &  3.5  \\
	    \\
	    1212  & 21\,500 &   3.78    &  307   &  8.0 & 24.0 &  8.0 & 24.0 &  8.0 & 24.0  \\
	    0079  & 29\,500 &   4.37    &  308   & 13.4 &  8.0 & 13.4 &  7.8 & 13.4 &  7.9  \\  
				\hline				
				\hline
			\end{tabular}

	\end{center}
\end{table*}

\section{Discussion} \label{s_discuss}

We now discuss our estimates of the stellar parameters and projected rotational velocities obtained in Sect. \ref{s_analysis}. We generally adopt our FT estimates of the latter but also consider the PF estimates when they might lead to different conclusions.

\subsection{Stellar masses and ages} \label{d_stellar}

As discussed by \citet{ind11}, the SMC has experienced significant recent star formation with peaks at 0-10\,Myr and 50-60\,Myr. Unlike the LMC, where this activity has been concentrated in the north and north-eastern regions, recent star formation in the SMC has not been found to have significant spatial structure. NGC\,346 as a young association is located in the central part of the brightest SMC \ion{H}{ii} region, N66 \citep{hen56}; its structure has been discussed by, for example, \citet{gou08} and \citet{hen08}. 

\begin{figure}
	\centering
	\includegraphics[scale=0.5]{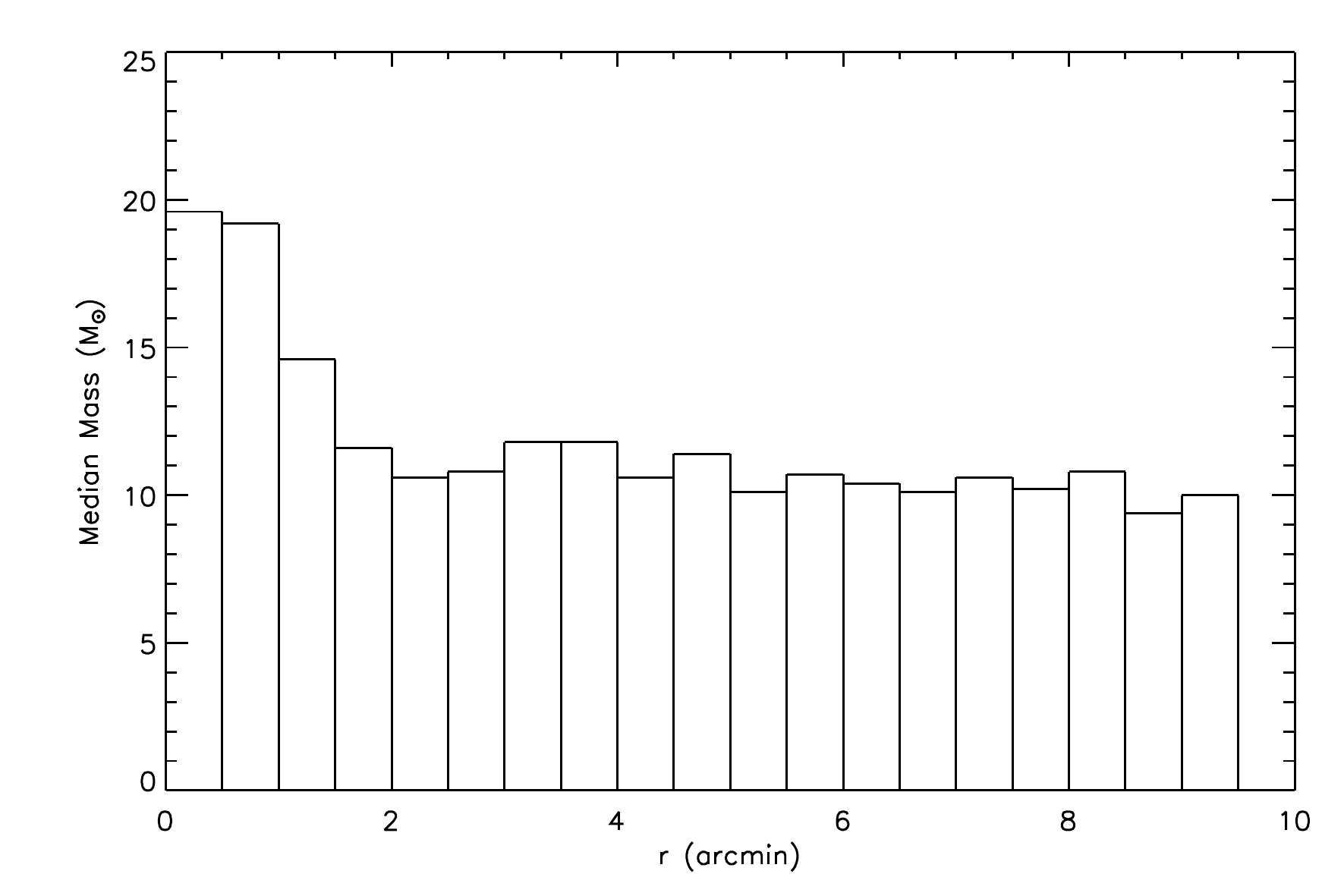}\\
	\includegraphics[scale=0.5]{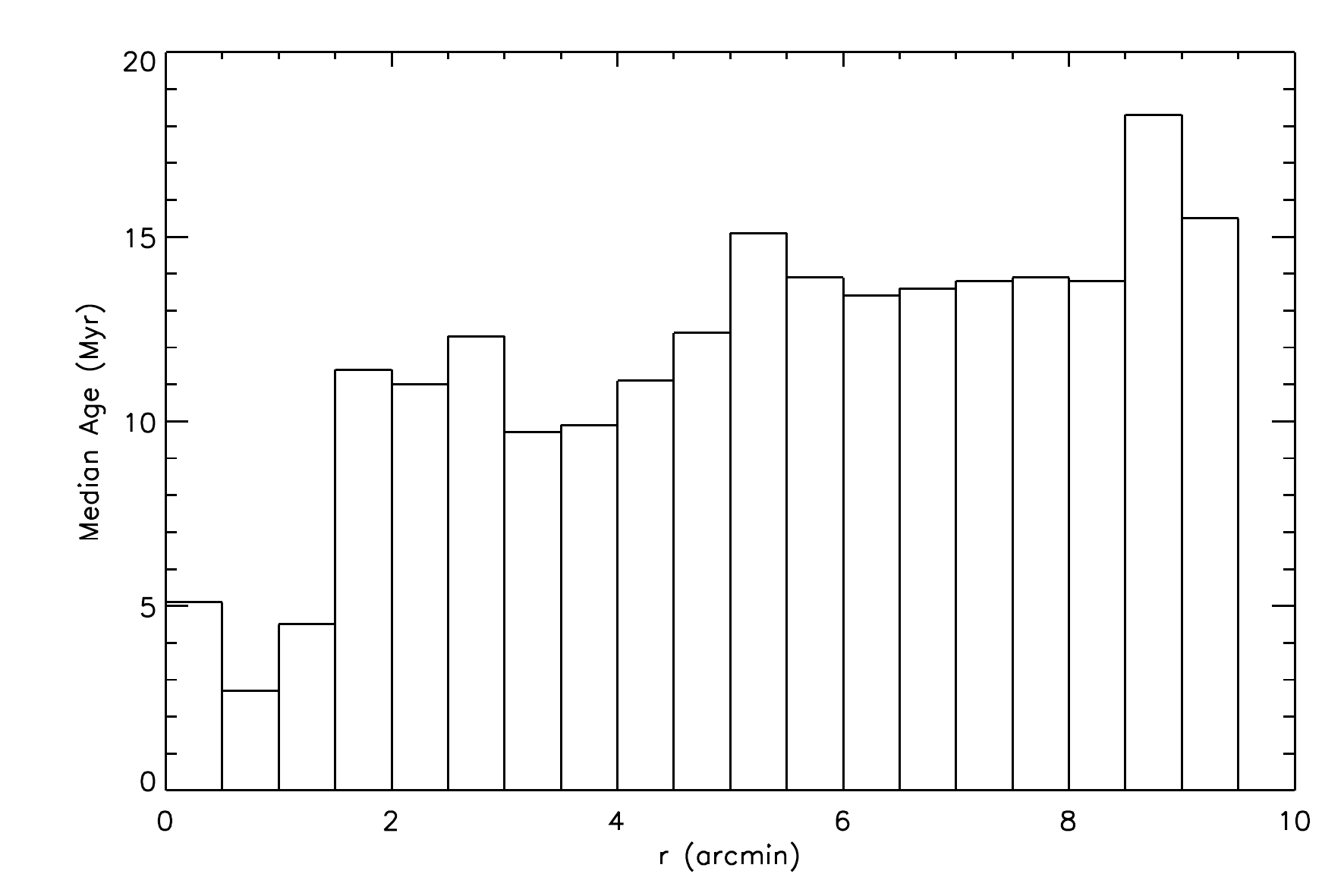}
	\caption[]{Median stellar mass (top) and age (bottom) distributions for our NGC\,346 spectroscopic
		sample as a function of radial distance from \#1001.}
	\label{f_radial_mass}
\end{figure}

Deep imaging from the {\em HST} has been used by \citet{sab08} to investigate the spatial variation of the present day mass function in the central part of NGC\,346 (with an outer radius of approximately 20\,pc, equivalent to $\sim$1\farcm13).  They found a steeper mass function with increasing radius, implying mass segregation for the most massive stars.  The medians of the estimated stellar masses and ages (see Table \ref{t_346results}) for our spectroscopic sample are shown as a function of radius ($r$) from the centre of NGC\,346 in Fig.~\ref{f_radial_mass}. Our adopted centre was 12\arcsec\  from that determined by \citeauthor{sab08}, from a consideration of source counts but this will not impact significantly on the discussion below.

Stars with radial distances, $r$\,$\la$\,2\farcm0, appear to be more massive, with a relatively flat distribution at larger distances. Additionally, these targets have smaller evolutionary age estimates. Kolmogorov--Smirnov (K--S) tests \citep{fas87} for both the estimated masses and ages return very low ($<$0.1\%) probabilities that the
samples from the inner and outer regions originated from the same parent population. 
Indeed, the mass distribution suggests that the stars are not bound to
the central cluster, in agreement with the definition of NGC\,346 by
\citet{gie10} as an `association', and reflecting the arguments by
\citet{ghk14} that it is comprised of two distinct components (with an
extended distribution of stars that formed hierarchically in addition
to the centrally-condensed cluster).

This decrease in the stellar mass and increase in the age estimates with radial distance mirrors the results of \citet{sab08} for the inner region of NGC\,346. It is also consistent with previous studies of the young stellar populations in our Galaxy and the LMC. For the former, \citet{get14a, get14b} identified similar trends for pre-main sequence (PMS) stars in the Orion molecular cloud complex, whilst recently \citet{get18} found that 80\% of a sample of 19 young clusters exhibited radial age gradients. For the latter, \citet{sch18a} deduced ages and masses for approximately 450 apparently single early-type stars in 30 Doradus observed by the VFTS. The medians for the ages and masses as a function of increasing distance from the central R136 star cluster show an increase and decrease respectively, in agreement with our results and those for the Galactic studies. Hence there is increasing evidence that such behaviour may be an ubiquitous feature of massive star formation. 

\begin{figure}
	\begin{center}
		\includegraphics[height=60mm]{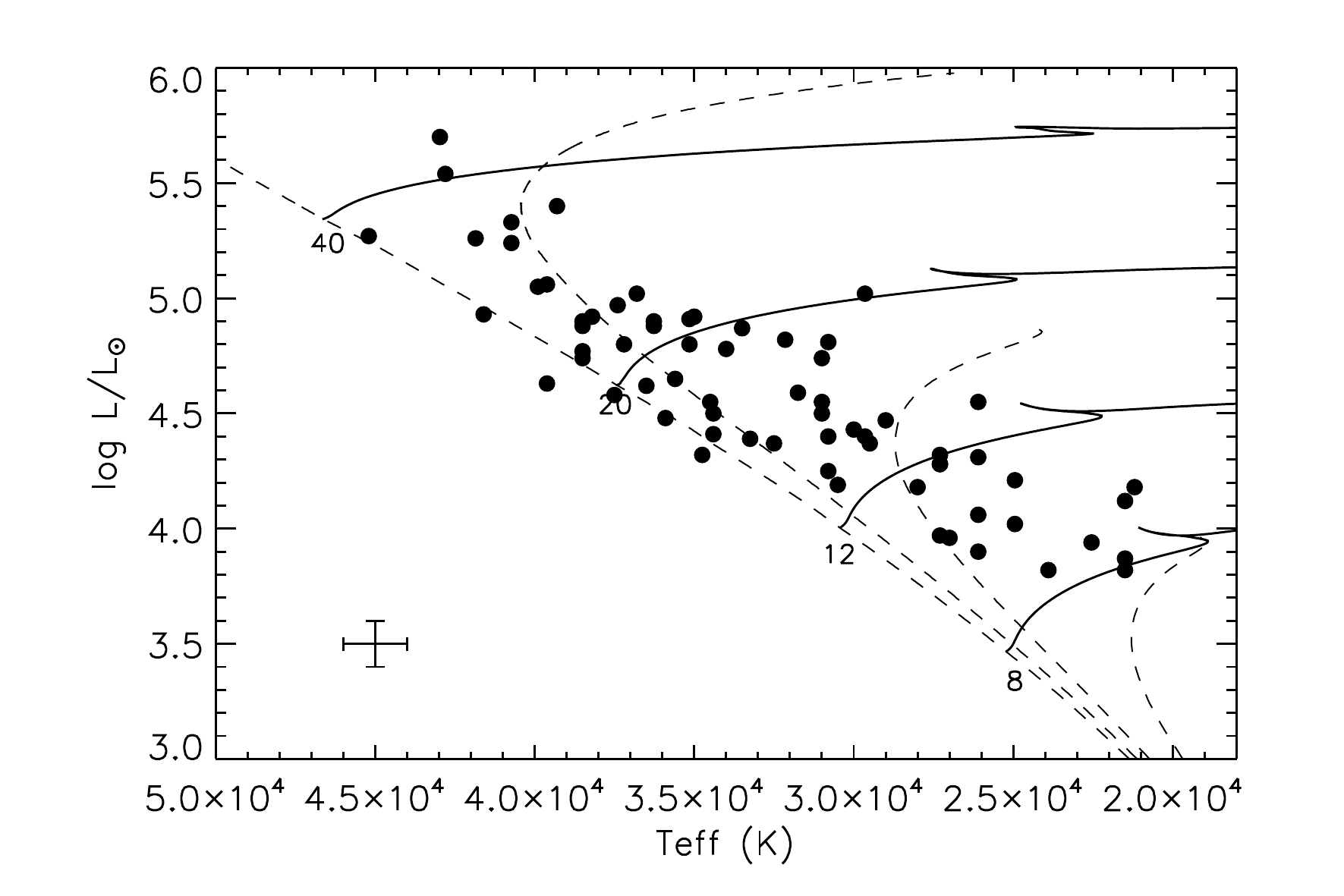}
		\includegraphics[height=60mm]{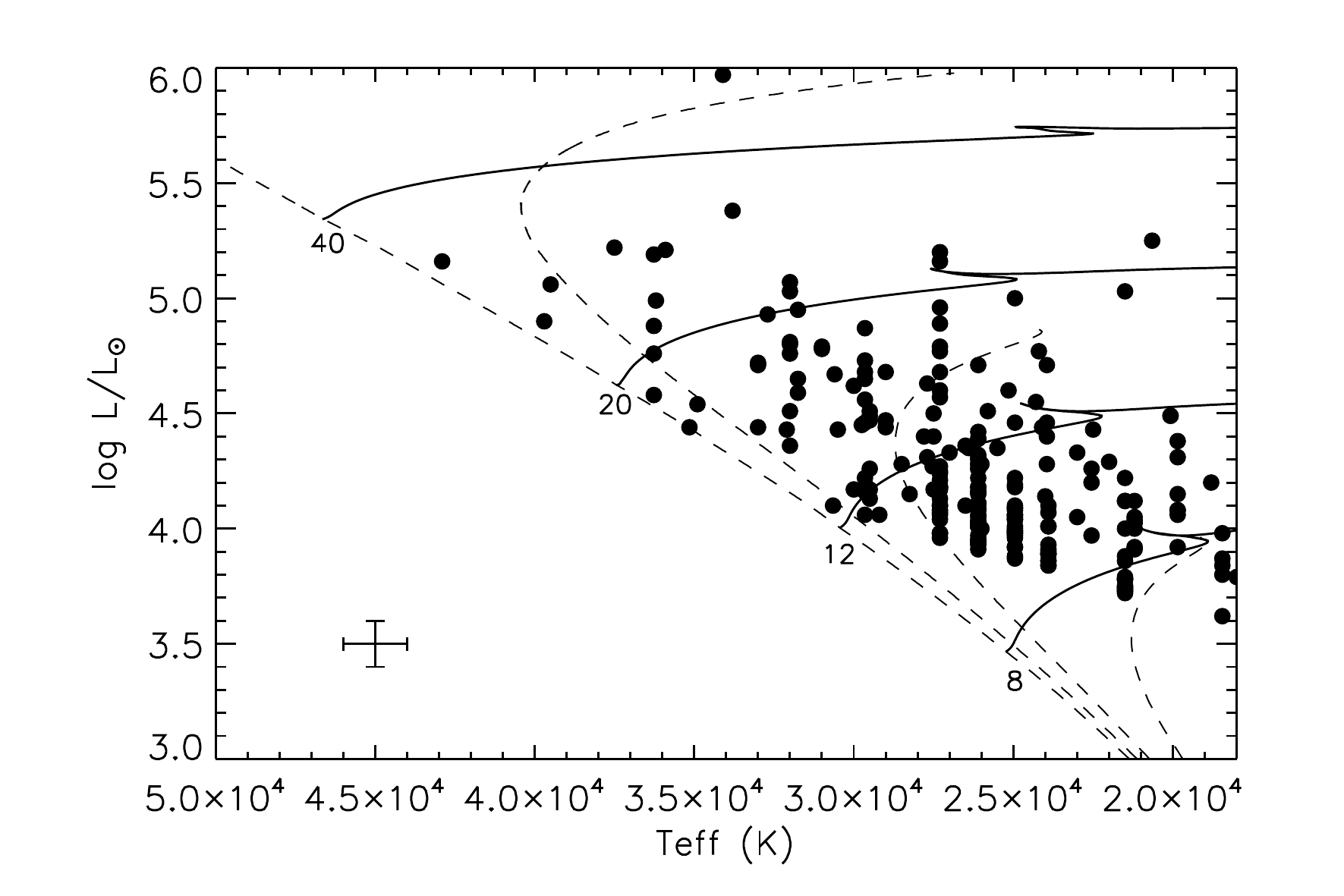}
		\caption[]{Hertzsprung--Russell diagrams of the inner ($r$\,$\leq$\,2\farcm0, upper panel)
			and outer ($r$\,$>$\,2\farcm0, lower panel) regions from our combined
			NGC\,346 sample.  Evolutionary tracks are for an initial rotational velocity of approximately 180\,\kms.  Isochrones (dotted lines) are for, from
			left-to-right: 0, 3, 10 and 30\,Myr.}
		\label{f_hrcluster}
	\end{center}
\end{figure}

Adopting the threshold of $r$\,$<$\,2\farcm0 implied by Fig.\ \ref{f_radial_mass} \citep[rather than the $r$\,$\simeq$\,3\farcm5 from][]{rel02}, Hertzsprung--Russell (H--R) diagrams for the inner and outer regions are shown in Fig.~\ref{f_hrcluster}. Evolutionary tracks and isochrones are from the SMC grid of \citet{bro11a} with initial rotational velocities of 180\,\kms, which is consistent with the median NGC\,346 projected rotational velocity of 136\,\kms\ and a random inclination of axes. However for targets near to the main sequence, the evolutionary tracks and isochrones are relatively insensitive to the choice of initial rotational velocity as can be seen from Figs. 5 and 7 of \citet{bro11a}. Our choice of model grid maintains consistency with the masses and ages estimated using {\sc bonnsai}. SMC evolutionary models generated with the Geneva evolutionary code \citep{geo13} have qualitatively similar tracks and isochrones but are only available for two initial rotational velocities. Hence we have not tried to use them to estimate masses and ages, although \citet{duf18} found that for VFTS targets with low \vsini, the Geneva tracks with zero rotation \citep{geo13a} yielded masses and ages that were consistent with the {\sc bonnsai} estimates within the observational uncertainties. 

A significant number of targets lie above the ZAMS and imply a range of stellar ages (cf. the isochrones in Fig.~\ref{f_hrcluster}) as discussed in Sect.~\ref{s_mass}.  Nonetheless, there may be other causes, for example some of these may be either detected (SB1 or SB2) or undetected binaries with the luminosity of the primary being overestimated. For a binary containing targets with a flux ratio of unity, this overestimate would be 0.3 dex, which would constitute an upper bound. Inspection of Fig.\ \ref{f_hrcluster} shows that particularly in the outer region a decrease of even 0.3\,dex in the luminosity would not bring most of the targets to the ZAMS. Previous investigations of young Galactic clusters \citep[see, for example,][]{str05, duf06} have also found targets above the main sequence, despite a better discrimination against binarity. Additionally it is possible that some of the targets may be pre-main sequence. However in that case it would be expected that more overluminous stars would be present in the inner region, whilst if anything the reverse would appear to be the case.

Thirteen targets (twelve classified as O-type and one as B0V) have {\sc bonnsai} age estimates of less than 1 Myr and 1-$\sigma$\ upper uncertainties ranging from 0.6 to 2.3\,Myr. Combining these age estimates and uncertainties to find 1-$\sigma$\ {\em upper limits} on the ages leads to a range of 1.0-2.6\,Myr with a median of 1.9\,Myr. Hence these massive stars (mainly situated in the inner region) argue for an age of less than 2\,Myr for the more recent period of star formation \citep[in agreement with the estimate from][based on four stars close to the zero-age main sequence]{wal00}. 

However it would appear that even the inner region of NGC\,346 is not a simple, coeval stellar population -- consistent with VFTS investigations of 30 Doradus \citep{sch18a}. Stars in this region (upper panel of Fig.\ \ref{f_hrcluster}) straddle several isochrones and it is clear that the cluster is more complex than just a simple single population \citep[e.g.][]{cig10,ghk14}.

Inspection of the targets with the largest age estimates (30-42\,Myr) shows that they are all classified as B3\,III. The corresponding dwarf B3\,V population, which would be expected to be younger, lies near the magnitude of the observational cutoff (see Sect.\ \ref{s_obs}) and may not have been well sampled. However Fig.~\ref{f_radial_mass} confirms that stars in the outer region of NGC\,346 generally appear older. Note that the intermediate-age cluster BS\,90 \citep{bic95} is only a couple of arcminutes to the north of NGC\,346, but is sufficiently old \citep[4.5\,Gyr,][]{sab07, roc07} that it does not contaminate our spectroscopic sample.

\begin{figure}
	\begin{center}
		\epsfig{file=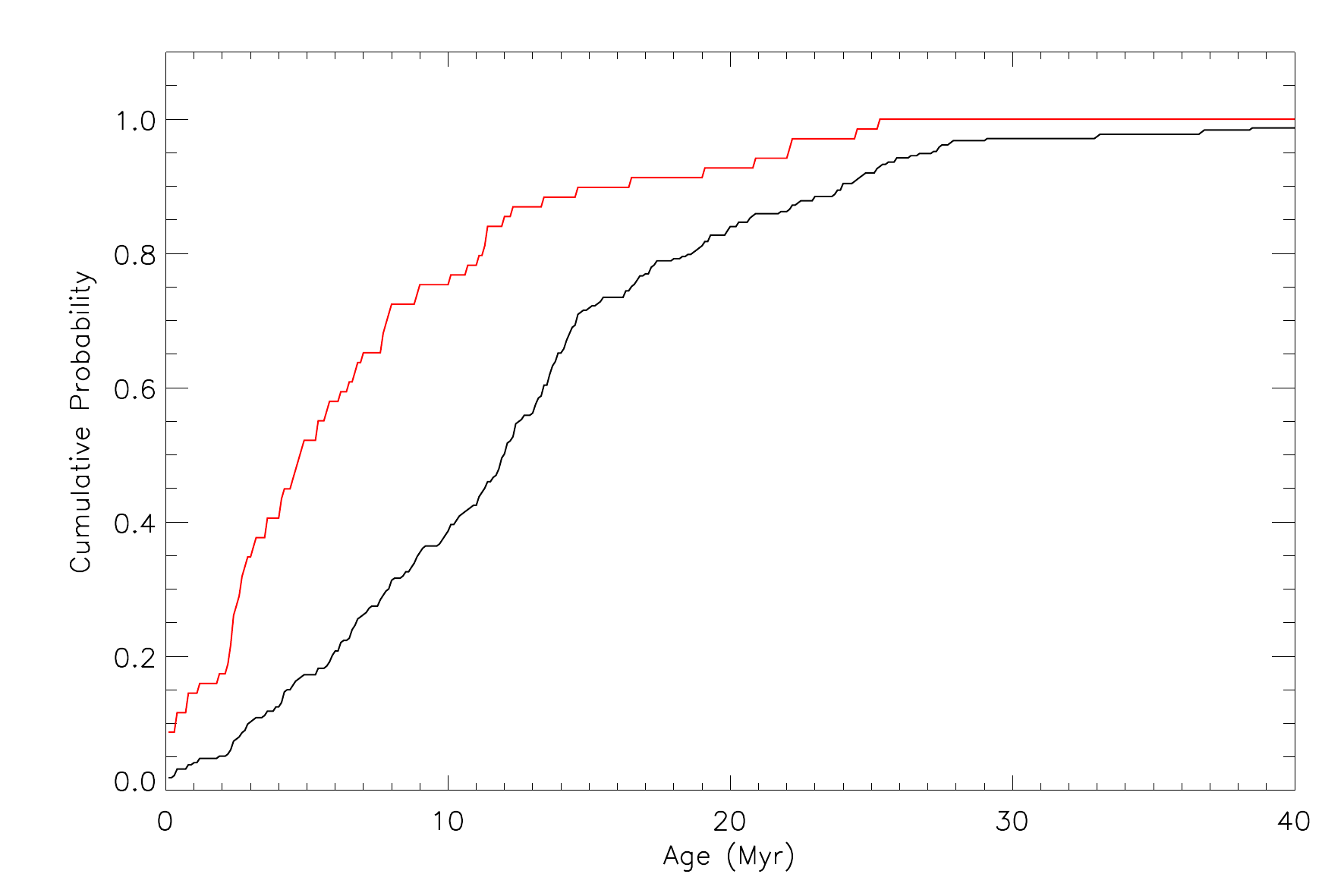, height=60mm, angle=0}
		\caption[]{Cumulative probability function for all 313 targets in NGC\,346 with age estimates (solid black line) and for the 69 targets with radial distances of less than 2\arcmin\ (red line).}
		\label{f_cdf_age}
	\end{center}
\end{figure}

In Fig.\ \ref{f_cdf_age}, the cumulative probability distribution (CPD) is shown for all the NGC\,346 age estimates. It is consistent with  multiple generations or, effectively, continuous star formation, as discussed by \citet{bro11b} for the FLAMES fields observed in the LMC. A histogram of the age distribution shows significant star formation for the last 28\,Myr peaking at an age of approximately 16 \,Myr.  As discussed above, our sample will be a mixture of cluster members and field stars. We have therefore repeated the above procedures for the 69 targets within a 2\arcmin\ radius of our adopted centre of NGC\,346. As can be seen in Fig. \ref{f_cdf_age}, these targets are (as expected) younger and imply significant star formation in the last 12\,Myr peaking in the last 4\,Myr. As this sample may also contain older field stars, this points to NGC\,346 being dominated by a very young stellar population.

\subsection{O-type stars outside the central cluster}
Three of the young O-type targets (\#0046, \#1114 and \#1144) lie outside the inner region at radial distances between 3\farcm8 and 6\farcm2. We have used the {\em Gaia} DR2 data release \citep{gaia1, gaia2} to search for pecularities in their proper motions (see Table \ref{t_GAIA}) consistent with them having been formed in the inner region. For a 10\arcmin\ radius centred on NGC\,346, there are approximately 1700 targets with $G$\,$\leq$\,16 and with proper motion estimates. These will be a combination of foreground and SMC targets but inspection of the proper motion distributions imply that they are heavily weighted to the latter. The median values of this sample are $\mu_{\rm RA}$\,$=$\,0.80\,\pmg\ and $\mu_{\rm Dec}$\,$=$\,$-$1.22\,\pmg.

In only one case (the $\mu_{\rm RA}$ of \#1114) is there a difference from the median value that is
significant at a 2$\sigma$-level. This star also has the largest
radial distance of the three, lying to the south east of the centre at
a separation of $\delta_{\rm RA}$\,$=$\,5\farcm67 and $\delta_{\rm
  Dec}$\,$=$\,2\farcm52. Assuming it was ejected from the centre of
NGC\,346 soon after birth, its estimated age of 0.4\,Myr would lead to
a peculiar proper motion of 0.85\,\pmg. This is substantially larger than its
proper motion relative to NGC\,346, i.e. $\delta(\mu_{\rm
  RA})$\,$=$\,0.25\,\pmg. However, the age of this object is
relatively poorly constrained with a 1-$\sigma$\ upper uncertainty of
1.9\,Myr. Increasing the age to 2.3\,Myr would then reduce the required
$\delta(\mu_{\rm RA})$ to 0.15\,\pmg, consistent with the {\em Gaia}
values. The separation in declination implies $\delta(\mu_{\rm Dec})$\,of 
$-$0.38 and $-$0.07\,\pmg\ for the two ages, the latter being
consistent with the effectively zero $\delta(\mu_{\rm Dec})$ implied by
the {\em Gaia} estimates.

For \#0046, the {\em Gaia} estimates imply effectively zero proper
motion with respect to NGC\,346. The estimated age (1.0\,Myr) and
upper uncertainty (0.9\,Myr) lead to predicted (relative) proper
motions as small as $\delta(\mu_{\rm RA})$\,$=$\,0.07\,\pmg\ and
$\delta(\mu_{\rm Dec})$\,$=$\,$-$0.10\,\pmg. These are consistent with
the {\em Gaia} estimates within the observational uncertainties,
particularly when considering the simple method used to ascertain the
motion of NGC\,346. Similar arguments apply to the proper motion
estimates for \#1144. In summary, the proper motions and age estimates
for all three stars are consistent with them having been formed close
to the centre of NGC\,346, although the uncertainties do not preclude
them having been formed elsewhere.

\begin{table}
{\small
	\centering
	\caption[]{Proper motions (\pmg) and radial distances ($r$) from the centre of NGC\,346 for selected targets from the DR2 {\em Gaia} data release. Also listed are the median values for targets with $r$\,$<$\,10$'$ from the centre of NGC\,346 and with $G$\,$\leq$\,16\,mag.}
	\label{t_GAIA}
	\begin{tabular}{llccll}
          \hline \hline
          Target  & ST  & \multicolumn{2}{c}{$r$} &\multicolumn{2}{c}{Proper motions ({\em Gaia})} \\
          & & [$'$] & [pc]     & ~~~~$\mu_{\rm RA}$ & ~~~~~~$\mu_{\rm Dec}$               \\
\hline
          0046  & O7 Vn    & 3.8  & $\phantom{1}$66.9 & 0.80$\pm$0.08  & $-$1.22$\pm$0.06 \\
          1114  & O9 V     & 6.2  & 109.2 & 1.05$\pm$0.10  & $-$1.22$\pm$0.07 \\
          1144  & O9.5 V   & 4.0  & $\phantom{1}$70.4 & 1.01$\pm$0.14  & $-$1.25$\pm$0.08 \\
          \\
          1134  & B0:e     & 2.5  & $\phantom{1}$44.0 & 0.82$\pm$0.09  & $-$1.34$\pm$0.06 \\
          1174  & B1-3e    & 6.5  & 114.5 & 0.97$\pm$0.10  & $-$1.25$\pm$0.08 \\
          \\
          \multicolumn{2}{l}{Median~~~~--}  & $<$10.0 & $<$176.1 & 0.80           & $-$1.22          \\
          \hline
		
		\hline
	\end{tabular}}
\end{table}

\subsection{Stellar rotational velocities}

\subsubsection{General properties}\label{d_vsini_general}

Projected rotational estimates have been obtained for effectively all our targets classified earlier than B9 with only seven SB2 systems, the one SB3 system and the peculiar target \#1024 (see Sect.\ \ref{1024}) having no estimates. The sample is predominantly B-type with 275 estimates compared with 46 O-type values. Additionally there are 21 OB-type supergiants (luminosity classes I-II) with 16 \vsini\ estimates. Significant macroturbulence has been found previously in O-type and early B-type supergiants \citep{rya02, sim14, sim17}. This would lead to our PF estimates (that include all excess broadening) being larger than the FT estimates (only the rotational component) and indeed a difference of 11$\pm$9\,\kms\ is found. 

The estimates cover a wide range of projected rotational velocities from less than 40\,\kms\ to approximately 380\,\kms. Dividing the sample into 40\,\kms\ bins, the most populous is that with 0$<$\vsini$\leq$40\,\kms, which implies that the sample contains a significant number of slowly rotating stars. This is confirmed by the deconvolution of the projected rotational velocity distribution discussed in Sect. \ref{d_ve}. Additionally there are two very rapidly rotating targets, \#1134 and \#1174, which are discussed further in Sect.\ \ref{d_1134_1174}.

\subsubsection{Mass variations} \label{d_vsini_mass}

\citet{hun08b} found lower projected rotational velocities for apparently single stars in the FSMS sample with $M$\,$>$\,25\,$M_{\sun}$ compared to those with lower masses in both the Magellanic Clouds and our Galaxy.  This was true even in the low metallicity SMC, where the effects of line-driven winds are expected to be weaker \citep{mok07}, although their high-mass sample was limited to only six stars.  

Our sample of non-supergiant targets that have mass estimates  $M$\,$\geq$\,25\,$M_{\sun}$ is over a factor of two larger with 14 objects. Again these targets have a lower median\footnote{It was not possible to calculate means given the significant number of targets with \vsini$\la$40\,\kms\label{median}.} FT \vsini\ estimate of 72\,\kms\ compared with 137\,\kms\ for the remaining targets; for the PF estimates the medians are 61 and 144\,\kms\ respectively. K--S tests (setting all the upper limits of the \vsini\ estimates to 40\,\kms) returned probabilities of 8.4\% (PF estimates) and 12.3\%, implying that differences in the projected rotational velocities of the two samples do not have a high level of statistical signifigance.

Projected rotational velocities have also been determined for both the O-type \citep{ram13, ram15} and B-type \citep{duf12, gar17} stars in the VFTS of 30~Doradus; unfortunately mass estimates are not available for all the targets. However the LMC models of \citet{bro11a} imply that 25\Msun\ models should have a ZAMS effective temperature of 38000--39500\,K, depending on the rotational velocity. In turn, the analysis by \citet{ram17} of the VFTS spectroscopy implies that this should correspond to a spectral type of O6.5 for luminosity class III-V objects. Hence we have repeated the above analysis for the apparently single non-supergiant VFTS targets and find median \vsini\ estimates of 111\,\kms\ for the 55 early-O type targets  cf. 173\,\kms\ for the remaining 406 lower-mass objects. Although these are higher than the medians for the NGC\,346 targets, they show the same qualitative behaviour. Additionally a K--S test returns a very low ($<$0.1\%) probability that the projected rotational velocities for two 30~Doradus samples originated from the same parent population. 

Further evidence for such a difference is provided by the estimated {\em rotational velocity} distributions for the apparently single non-supergiant VFTS O-type \citep{ram13} and B-type \citep{duf12} samples. The former appears unimodal with a mode at approximately 100\,\kms\ and an extended higher velocity tail, while the latter is bi-modal with maxima at 60 and 300\,\kms. In turn this leads to median rotational velocities of 160\,\kms\ and 250\,\kms, respectively. The LMC models of \citet{bro11a} imply that the boundary between the two samples will be at a mass of 16\,\Msun. Although this boundary differs from that considered above of 25\Msun, the medians again indicate that the more massive stars in general have lower rotational velocities.

In summary there is evidence that in the LMC, targets with masses, $M$\,$\geq$\,25\,$M_{\sun}$, have lower median projected rotational velocities than less massive early-type stars. A similar difference is seen in our results for the smaller sample in NGC\,346, but it is not as statistically significant.

\subsubsection{Spatial variations}\label{d_vsini_spatial}

B-type stars in Galactic clusters \citep[see, for example][]{hua06, wol07} appear to rotate more quickly than those in the field, with some evidence in support of this in the LMC \citep{kel04}.  In Sect.\ \ref{d_stellar}, targets within 2\farcm0 of the cluster centre were found to have a higher median mass and a smaller median age than the rest of the sample. We have therefore calculated medians for the \vsini\ estimates of the apparently single B-type stars (excluding luminosity class I-II supergiants) in these two spatial regions and summarize the results in Table \ref{t_medians}.\footref{median}

\begin{table}
	\centering
	\caption[]{Medians of \vsini\ estimates of single non-supergiant B-type stars for the inner and outer regions of NGC\,346 obtained by the profile fitting (PF) and Fourier transform (FT) methods, together with the number of stars in each sample ($N$).}
	\label{t_medians}
	\begin{tabular}{lrrr}
		\hline \hline
		Region                        & $N$ & \multicolumn{2}{c}{Medians} \\
		&     & PF & FT                               \\
		\hline
		NGC\,346 ($r$\,$<$\,2\farcm0) & ~33 &  187 & 195 \\
		NGC\,346 ($r$\,$>$\,2\farcm0) & 187 &  143 & 137 \\
		NGC\,330                      & ~72 &  143 & --\\  		
		30~Doradus                        & 250 &  --  & 195 \\
		NGC\,6611                     & 24  &  134 & -- \\
		 
				\hline
	\end{tabular}
\tablefoot{Results for NGC\,330 and 30~Doradus are median values for the `field-like' samples from \citet{hun08a}
and \citet{duf12}, respectively. Results for the Galactic cluster NGC\,6611
are from \citet{duf06}.}
\end{table}

In Table \ref{t_medians}, we also list the medians from the FSMS \vsini\ estimates of \citet{hun08a} toward the SMC cluster, NGC\,330 (adopting the same selection criteria as for NGC\,346). These were obtained using a PF methodology similar to that adopted here and provide a predominantly SMC ‘field-like’ sample as the large majority lie well beyond the cluster radius \citep[see][]{eva06}. We find that the median projected rotational velocities in the outer region of NGC\,346 and in the (predominantly field sample) of NGC\,330 are in good agreement. Although both samples are almost certainly not purely field stars (and NGC\,330 is older than NGC 346), these results serve to illustrate the slower rotational velocities compared to the inner region of NGC\,346 in agreement with the Galactic studies. 

The stars at larger radial distances have on average larger ages than those within the inner 2\farcm0 of the cluster as discussed in Sect.\ \ref{d_stellar}. Hence the variation in median projected rotational velocity may reflect the stellar rotation decreasing as the stars evolve during their hydrogen core burning phase. Inspection of the SMC models of B-type stars by \citet{bro11a} shows that this can occur with rotational velocities decreasing by up to 20\%. However this is limited to models with low masses ($<12$\Msun) near to the end of this evolutionary phase. For example, for a gravity, \logg$\sim$3.8\,dex (which is less than the median gravity for our B-type sample of 3.95\,dex) changes are less than 5\% with the rotational velocity having increased in some cases. Hence environmental effects such as those discussed by \citet{wol07} could also be important.

However these results must be treated with some caution. K--S tests (using the same procedure for \vsini\ upper limits as in Sect.\ \ref{d_vsini_mass}) for the samples in inner and outer regions of NGC\,346 returned P-values of 16\% (PF estimates) and 20\% (FT estimates). Tests using samples for the inner region of NGC\,346 and NGC\,330 also returned high probabilities that they originated from the same populations. Hence we conclude that our SMC results are consistent with the Galactic investigations, although our samples remain too small to have a high level of statistical signifigance.

\subsubsection{The rapidly rotating targets, \#1134 and \#1174} \label{d_1134_1174}

Two targets, \#1134 and \#1174, have estimates of their projected rotational velocity  from the \ion{He}{i} line at 4026\,\AA\ in excess of 500\,\kms\ (see Table \ref{t_346results}). As discussed by, for example, \citet{duf11}, it is difficult to reliably estimate such large values due to the line profiles becoming very broad and shallow. We have therefore also obtained FT estimates from the other diffuse \ion{He}{i} lines at 4143, 4387 and 4471\,\AA. These lead to mean values of 519$\pm$15 (\#1134) and 508$\pm$17 (\#1174) in reasonable agreement with the estimates from the 4026\,\AA\ line. 

Irrespective of the precise values, it is clear that these stars have significantly greater projected rotational velocities than the rest of the sample. Adopting the effective temperatures, luminosities and masses estimated in Table \ref{t_346results} and the SMC grid of evolutionary models of \citet{bro11a} leads to estimates of the critical velocity of 736\,\kms\ (\#1134) and 724\,\kms\ (\#1174), assuming an initial equatorial rotational velocity, \vi$\simeq$530\,\kms. These critical velocities are relatively insensitive to the initial rotational velocity with a change of less than 10\,\kms\ for the models with \vi$\simeq$590\,\kms. In turn, this implies that both stars are rotating at more than 70\% of the critical velocity. Indeed, given that the average values for $\sin i$\ assuming a random distribution of inclination axes is 0.785 \citep{gra05}, it is possible that both stars are rotating at near critical velocities.

\citet{duf11} previously discussed an LMC star, VFTS\,102 with a spectral type O9: Vnnne and a projected rotational velocity, \vsini\,=\,600$\pm$100\,\kms, lying approximately 12 pc from the X-ray pulsar PSR J0537$-$6910 in the plane of the sky. They suggested that this object originated from a binary system with its high projected rotational velocity resulting from mass transfer from the progenitor of PSR J0537$-$691. We have therefore searched for pulsars and supernova remnants (SNRs) in the vicinity of \#1134 and \#1174. 

The closest pulsar in the ATNF Pulsar Catalogue\footnote{An updated version of this catalogue is available from: http://www.atnf.csiro.au/research/pulsar/psrcat} \citep{man05} is PSR J0100$-$7211, identified by \citet{lam02} as an anomalous X-ray pulsar. Subsequently, \citet{dur05} identified a {\em possible} optical counterpart from {\em HST} imaging but this was not confirmed from further deep imaging with the {\em HST}. Its X-ray properties were discussed by \citet{mcg05}, who found a characteristic age of ($P$/2$\dot{P}$) of 6800\,yr. Both \#1134 and \#1174 lie approximately 0.15$^\circ$ from the pulsar corresponding to approximately 160\,pc at the distance of the SMC. Adopting the characteristic age would then require a relative velocity in excess of 23\,000\,\kms\ for them to have previously been physically connected. Hence we conclude that it is highly unlikely that either star were physically associated with the progenitor of PSR J0100$-$7211. This is confirmed by their proper motion estimates (see Table \ref{t_GAIA}), which are consistent with the median values for NGC\,346. Similar arguments would apply to other SMC pulsars, although we note that their beamed emission implies that other such objects may lie closer to our two targets but remain undetected at present. 

The nearest catalogued supernova remnant to either target is B0057$-$72.2 (also designated B0057$-$72.4). The position found by \citet{fil02} from radio observations leads to angular separations of 3\farcm7 (\#1134) and 5\farcm7 (\#1174), which correspond to sky separations of approximately 65 and 100\,pc respectively. However these distances must be treated with caution given the large angular size of the SNR. For example, \citet{dav76} from optical imaging found major and minor axes of 14\arcmin\ and 11\arcmin\ with evidence for sub-structure.

Given the large rotational velocities combined with the Be nature of both targets (and nebular contamination in \#1134) it was difficult to recover robust estimates of their radial velocities. Nonetheless, estimates from the He~\1 lines are not inconsistent with what we would expect for the region \citep[cf. results from][]{eva06}, i.e. they do not appear to be significantly offset from the local systemic value, suggesting neither is a
significant (RV) runaway.

In summary, there is no direct evidence that either of these targets originated from a binary system where their companion became a supernova. However, we cannot discount that they were originally the secondary components of a binary system. Mass transfer from the primary would then lead to their rapid rotation and their characterisation as a single star as discussed by, for example, \citet{deM13, deM14} and \citet{bou18}. Although their evolutionary pathways remain unclear, their projected rotational velocities suggest that they may have had a different evolutionary history to the rest of our sample.

\subsubsection{Metallicity effects} \label{d_met}

Our projected rotational velocity estimates can be compared with those
found from the VFTS to search for effects due to the different
metallicities of the Clouds. The VFTS spectroscopy was obtained with
same instrumentation to that used here, while the data reduction and
analysis methodologies were similar. Therefore, we do not expect any
major systematic differences between estimates from the two datasets.

\begin{table*}
	\caption[]{Statistical tests on the estimated projected rotational velocities for main-sequence
(MS) stars in the NGC\,346 (346), VFTS (30~Dor) and NGC\,6611 (6611) samples. The first four columns identify the samples and the number of estimates. For the B-type samples, targets classified later than B3 were excluded. The latter four columns list the D-statistic and probability that the two samples come from the same parent population (P) for the Kolmogorov--Smirnov (K--S) and Kuiper tests.}\label{t_KS}
	\begin{center}
		\begin{tabular}{lccccccc}
			\hline\hline
			Sample			& \multicolumn{3}{c}{Number} 	& \multicolumn{2}{c}{K--S} 	& \multicolumn{2}{c}{Kuiper} \\
			&    346 & 30~Dor & 6611 & D & P & D & P \\
			\hline 
			Single MS O-type & 34  & 189 & -- & 0.117 & 0.804 & 0.159 & 0.936 \\
			Binary MS O-type & 9   & 104 & -- & 0.154 & 0.981 & 0.253 & 0.982 \\
			\\
			Single MS B-type & 218 & 289 & -- & 0.119 & 0.055 & 0.133 & 0.173 \\
			Binary MS B-type & 40  & 95  & -- & 0.203 & 0.175 & 0.272 & 0.178 \\
			\\	
			Single MS B-type & 218 & --  & 24 & 0.210 & 0.262 & 0.351 & 0.063 \\
			\hline
		\end{tabular}
	\end{center}
\end{table*}

\begin{figure}
	\centering
	\includegraphics[scale=0.5]{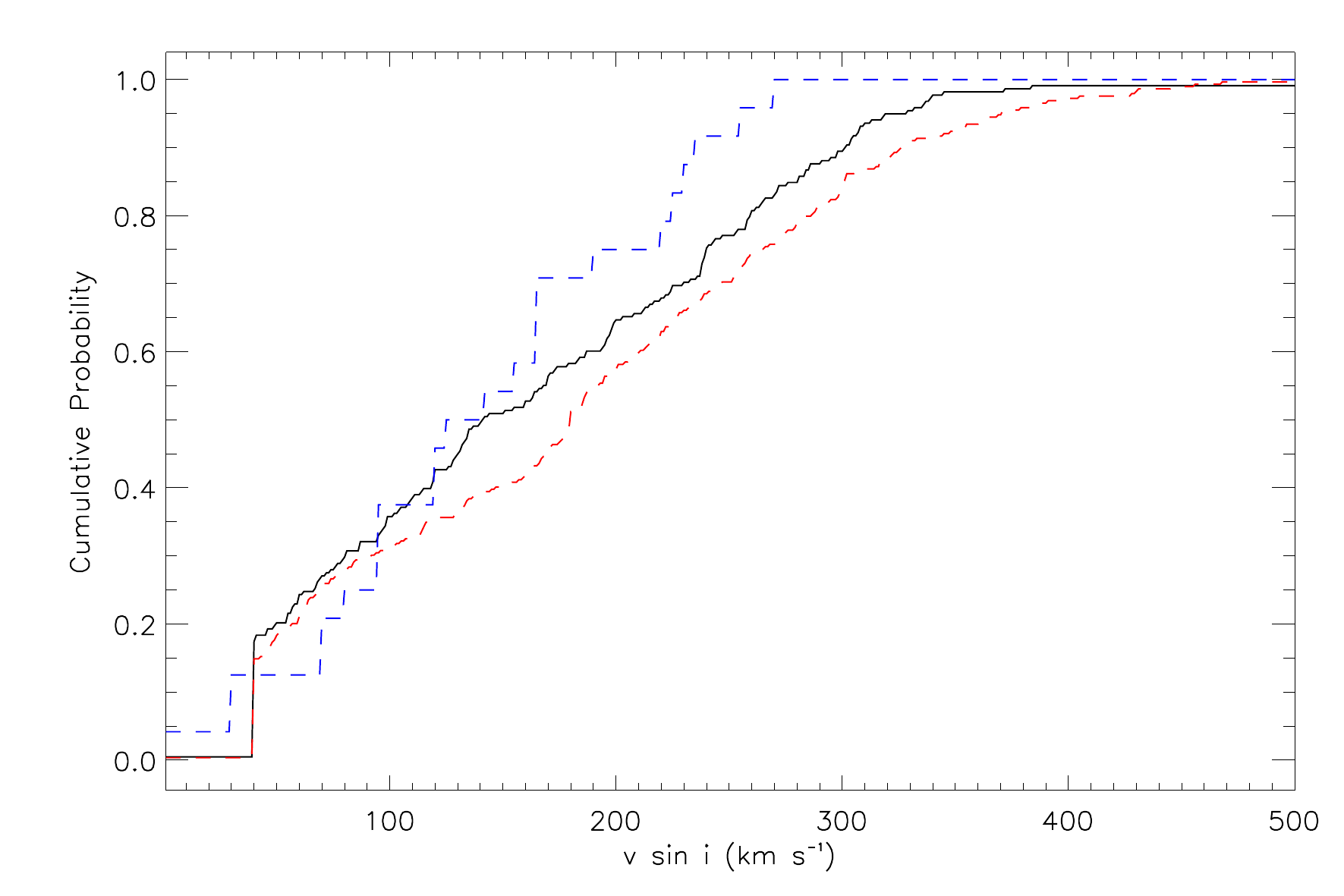}\\
	\caption[]{Cumulative probability distributions for the single B-type stars in NGC\,346 (black line), 30~Doradus (red dotted line), NGC\,6611 (blue dotted line). Upper limits on \vsini\ estimates have been set to 40\,\kms.}
	\label{f_CDF}
\end{figure}

Estimates for the O-type single and binary targets were obtained by \citet{ram13,ram15}, with those for the B-type samples being taken from \citet{duf12} and \citet{gar17}. We have limited our comparison to samples of hydrogen core burning objects (i.e. excluding luminosity classes I and II). In Figure \ref{f_CDF}, we display the CPD for our largest sub-sample of single B-type stars, together with those for the same cohort in 30~Doradus \citep{duf12}.

The distributions are similar although there is some evidence for the B-type stars in 30~Doradus having higher rotational velocities. In part, this may reflect a higher fraction of this sample being associated with clusters \citep[][]{eva11}. This would be consistent with the median projected rotational velocity (see Table \ref{t_medians}) for 30~Doradus being in good agreement with that for the inner region of NGC\,346 but larger than those for the field dominated sample.
 
We have investigated this further using statistical tests for the four sub-samples of the O- and B-type targets split into apparently single and binary systems. K--S and Kuiper tests were performed and the results are summarized in Table \ref{t_KS}. For all cases, there is no evidence that the projected rotational velocity distributions in NGC\,346 and 30~Doradus are different at a 5\% significance level. In turn this implies that there is no evidence for the rotational velocity distributions being different in the two metallicity environments.

For completeness, in Fig.\ \ref{f_CDF} we also show the CPD for the projected rotational velocity estimates of B-type stars in NGC\,6611 based on the FSMS observations \citep{duf06} and list the median value in Table \ref{t_medians}. The CPD implies that the NGC\,6611 targets have, on average, lower rotational rates, which is supported by its lower median. As the NGC\,6611 targets should be all cluster members, this would be consistent with faster rotation occurring in lower metallicity environments. However the statistical tests summarized in Table \ref{t_KS} show that these differences are not significant at a 5\% level, due to the relatively small sample size for NGC\,6611.
	
\subsubsection{Deconvolution of the \vsini\ distribution} \label{d_ve}

\begin{figure}
	\centering
	\includegraphics[scale=0.5]{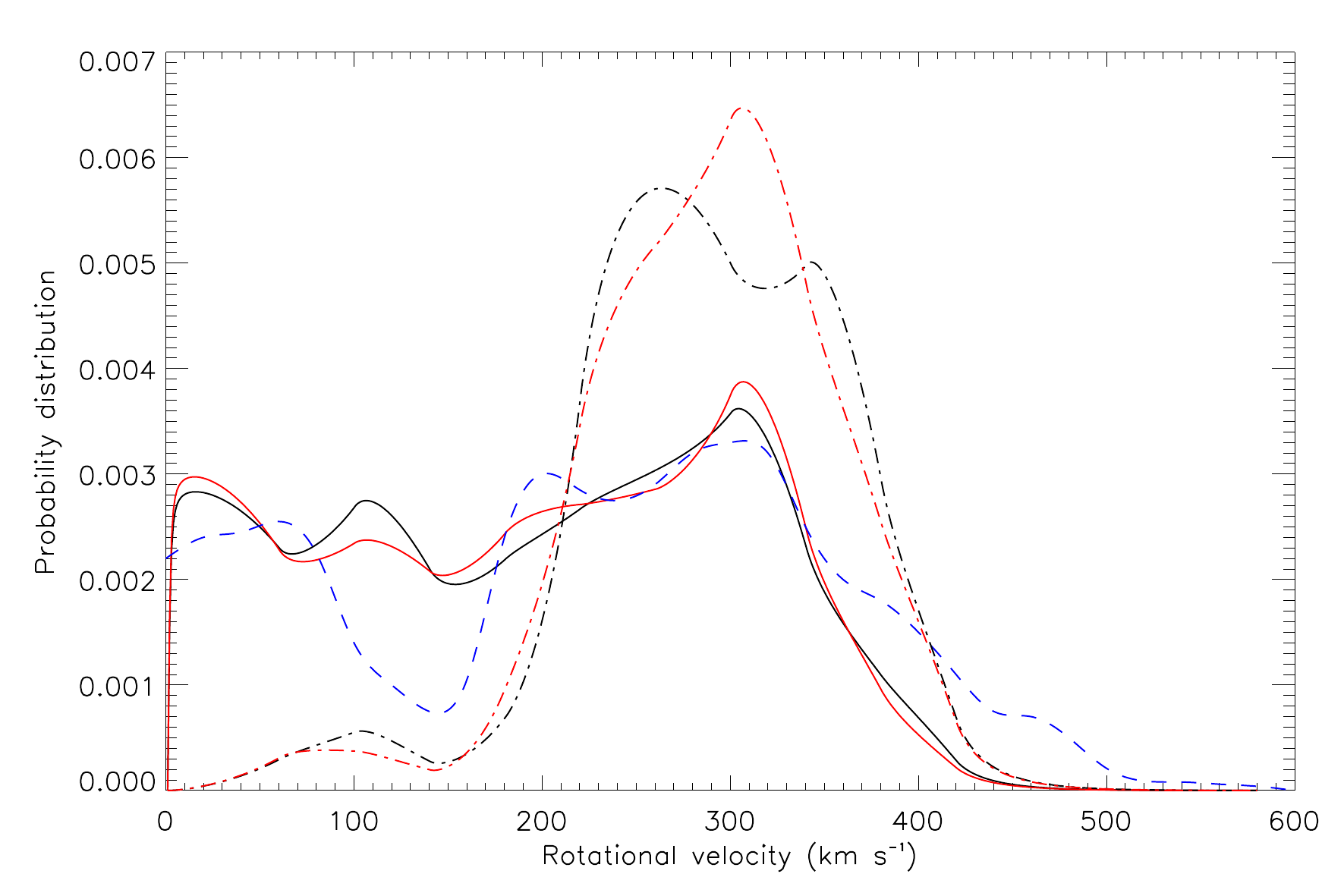}\\
	\caption[]{Rotational velocity probability distributions, $f(v_e)$, for the apparently single B-type stars in NGC\,346 deduced from the PF (black line) and FT (red line) estimates, and for the equivalent sample in 30~Doradus (dotted blue line) of \citet{duf12}. Also shown are the equivalent distributions for the apparently single Be-type (dash-dotted black and red lines).}
	\label{f_v_e}
\end{figure}
Our sample of apparently single non-supergiant B-type targets is large enough to estimate the current distribution of their rotational velocities. Assuming that their rotation axes are randomly distributed, we can infer the probability density function distribution, P(\ve), using the iterative procedure of \citet{luc74} as implemented by \citet{duf12}, where further details can be found. The two extremely fast rotators (\#1134 and \#1174) have been excluded from the deconvolution for numerical stability reasons.

The probability distributions implied by both the FT and PF \vsini\ estimates are shown in Fig. \ref{f_v_e}, together with that for the equivalent sample in the 30~Doradus region \citep{duf12}. The two probability distributions inferred for the NGC\,346 are broadly similar, reflecting the good agreement between the PF and FT estimates as discussed in Sect. \ref{s_vsini_method}. 

As discussed in Sect. \ref{d_met}, the CDFs for the \vsini\ estimates of NGC\,346 and 30~Doradus were consistent with them having the same parent population and this must be borne in mind when considering possible differences. In general, the distributions are similar with a significant population of stars having small rotational velocities and with a second maximum at approximately 300\,\kms. The NGC\,346 sample has some additional structure between 80 and 200\,\kms\ but given the differences between the two P(\ve) distributions based on the PF and FT estimates, this cannot be considered significant. Additionally, as discussed by \citet{duf12}, the use of a bin size of 40\,\kms\ in the de-convolution implies that structure on this scale or smaller is unlikely to be real. There may also be differences at high rotational velocities (\ve\,$>$\,350\,\kms) but given the relatively small number of targets with large projected rotational velocities (see Fig.\ \ref{f_CDF}), these can again not be considered significant.

For both regions approximately 10\% of the targets have rotational velocities of less than 40\,\kms. For 30~Doradus, \citet{duf18} found that a significant fraction of these targets had enhanced nitrogen abundances that were inconsistent with current single star evolutionary models. They considered possible explanations, of which the most promising would appear to be braking due to magnetic fields \citep{mor12,aer14a} or stellar mergers with subsequent magnetic braking \citep{sch16}. A similar investigation of the B-type stars in NGC\,346 would be valuable.

Our best estimates of the probability density function for the apparently single B-type stars (excluding supergiants) are given in Table~\ref{t_veq}. These values require a judgement about the nature of the smaller scale structure as discussed above. Also listed is the cumulative distribution function corresponding to these estimates and, because this is truncated at \ve$\leq 500$\,\kms, it does not reach unity. We recommend that either the range up to the critical velocity is populated or that the probability density function is renormalised, depending on the nature of the application. 

\begin{table}
	\caption{Estimates of the probability density of the rotational velocity, P(\ve), and its cumulative distribution function (cdf) for the NGC\,346 apparently single B-type sample. Supergiants (luminosity classes I and II) were excluded from the de-convolution.}
	\label{t_veq}
	\begin{center}
		\begin{tabular}{rccccll}
			\hline\hline
			\ve   & P(\ve)$\times 10^3$  & cdf   \\
			\hline
			0    &  2.90  &  0.0000  \\
			20   &  2.90  &  0.0580  \\
			40   &  2.75  &  0.1145  \\
			60   &  2.40  &  0.1660  \\
			80   &  2.25  &  0.2125  \\
			100  &  2.20  &  0.2570  \\
			120  &  2.15  &  0.3005  \\
			140  &  2.15  &  0.3435  \\
			160  &  2.20  &  0.3870  \\
			180  &  2.25  &  0.4315  \\
			200  &  2.45  &  0.4785  \\
			220  &  2.60  &  0.5290  \\
			240  &  2.80  &  0.5830  \\
			260  &  2.95  &  0.6405  \\
			280  &  3.10  &  0.7010  \\
			300  &  3.60  &  0.7680  \\
			320  &  3.50  &  0.8390  \\
			340  &  2.55  &  0.8995  \\
			360  &  1.70  &  0.9420  \\
			380  &  1.00  &  0.9690  \\
			400  &  0.60  &  0.9850  \\
			420  &  0.25  &  0.9935  \\
			440  &  0.10  &  0.9970  \\
			460  &  0.05  &  0.9985  \\
			480  &  0.02  &  0.9992  \\
			500  &  0.00  &  0.9994  \\			
			\hline
		\end{tabular}
	\end{center}
\end{table}

\subsubsection{Rotational velocities of the Be-type stars} \label{d_Be_ve}

Classical Be-type stars have prominent emission features in their Balmer line spectrum, indicating the presence of a geometrically flattened, circumstellar disc \citep{qui94, qui97}. \citet{zor97} have estimated that approximately 17\% of Galactic stars showed the Be phenomenon, with the highest fraction around  B1e\,–\,B2e. Emission is also often seen in the helium and iron spectra, with silicon and magnesium emission lines being seen in some stars \citet{por03}. A recent review of the Be phenomenum has been given by \citet{riv12a} 

Seventy three targets have been identified as Be-type equating to 25\% of the total B-type sample; this increases to 27\% if supergiants are excluded. Nine of the Be-type stars are binaries and excluding these leads to 29\% of the apparently single stars (excluding supergiants) being Be-type. These percentages are higher than those found by \citet[17\%]{zor97} for Galactic field stars and by \citet[18\%]{mar06a} towards NGC\,2004 in the LMC (this sample was predominantly field stars). Our Be-type percentages are similar to that of \citet[26\%]{mar07a} for the field towards NGC\,330 and \citet[27\%]{bon10} from an infrared investigations of the SMC. \citet{mar06a, mar07a} found tentative evidence that the proportion of Be-type stars in clusters might be higher but these were based on small numbers of B-type stars (LMC: 19 stars, SMC: 18 stars). For our sample of 33 apparently single B-type stars within 2\farcm0 of the centre of NGC\,346, we find 6 Be-type stars (excluding \#1100 which is possibly a Herbig B-star) leading to a percentage of 18\%, with an upper 95\% confidence limit of 33\%. Hence although there is no evidence for an increased Be-type fraction in the centre of NGC\,346, our sample is too small to exclude that possibility

The mechanisms underlying the Be phenomenon remain unclear but such stars in our Galaxy are believed to be fast rotators with velocities that range from $\sim$60\% up to 100\% \citep{cha01, fre05, cra05, riv06, eks08} of the critical limit \citep{tow04}. For our sample, this is consistent with the small number of Be-type stars with low \vsini\  estimates and a correspondingly large median value of 240\,\kms\ compared with those in Table \ref{t_medians}. For the apparently single Be-type stars, we have performed a de-convolution of the \vsini\ estimates using the same methodology as summarized in Sect.\ \ref{d_ve}. In Fig.\ \ref{f_v_e}, these are shown for both the PF and FT \vsini\ estimates. The sample size is small and hence caution should be exercised in interpreting these results. For example, there appears to be a population of slowly rotating Be-type stars (\ve$\sim$100\,\kms) but this is based on a relatively small number (9-10) of targets with correspondingly low \vsini\ estimates.

The rotational velocities of our Be-type sample mainly lie in the range $\sim$200--450\,\kms. \citet{dun11} inferred rotational velocity distributions for Be-type samples in the Magellanic Clouds assuming a Gaussian distribution. For the SMC, these peaked at rotational velocities of 260\,\kms\ and 310\,\kms\ for the FSMS and the survey of \citet{mar06} respectively in good agreement with the results found here. Adopting the critical velocity estimates from Sect.\ \ref{d_1134_1174} would then imply that they are rotating at $\sim$0.3--0.6 of this velocity. This is lower than that generally found for Galactic targets but as discussed by \citet{riv12a}, there may be systematic biases in our \vsini\ estimates.  These would include additional absorption in shell stars and line emission in Be stars, as well as the presence (of a presumably narrower lined) secondary.

In summary, the projected rotational velocities of our Be-type stars are systematic larger than those of the rest of the B-type stars. This implies rotational velocities between 200--450\,\kms, although the Be-type velocities may be subject to systematic biases.

\section{Conclusions}

We have presented spectral types and estimates of the atmospheric parameters and projected rotational velocities for the hot star population towards NGC\,346. Additionally, we have inferred masses and ages using the {\sc bonnsai} package \citep{sch14}. Our conculsions are:
\begin{enumerate}
	\item Targets towards the inner region of NGC\,346 have higher median masses and smaller ages than the rest of the sample. There appears to be a population of very young targets with ages of less than 2\,Myr. These are predominantly in the inner region, while the three young targets at greater radial distances could also have been formed nearer the centre.
	\item The more massive targets may have lower median projected rotational velocities consistent with previous studies of the Magellanic Clouds.
	\item Targets close to the centre of NGC\,346 have a higher mean projected rotational velocities than those at larger radial distances consistent with previous Galactic studies.
	\item Two targets (\#1134 and \#1174) have very large projected rotational velocities and are rotating in excess of 70\% of their critical velocities. Their origin is investigated and especially whether they could be the original secondaries in a binary where the primary evolved into a supernova. However it is not possible to draw firm conclusions from the current observations.
	\item A comparison with the 30~Doradus region in the LMC finds no evidence for significant differences in early-type stellar rotational velocities with metallicity. There is some evidence that the SMC targets rotate faster than those in young Galactic clusters but this is not statistically significant. These results agree with the conclusions from ultraviolet studies of early-type stars in the Clouds \citep{penny04,pg09}.
	\item The projected rotational velocities for the apparently single B-type hydrogen burning targets have been deconvolved to infer their rotational velocity distribution. This shows that a significant number have low rotational velocities ($\simeq$10\% with \ve$<$40\,\kms). Additionally the distribution peaks at a rotational velocity of approximately 300\,\kms.
	\item Relatively large projected rotational velocity have been found for our Be-type sample and imply rotational velocities between 200--450\,\kms. As our Be-type \vsini\ estimates may be systematically biased to lower values, this difference between the rotational velocities of the B-type and Be-type stars may be even larger.

\end{enumerate}

The results presented here are based on samples of approximately 350 and 750 targets in the SMC and LMC respectively. Further progress in understanding the rotational velocities of early-type stars in the Magellanic Clouds will require significantly larger samples of targets. Additionally it will be important to ensure that the samples are well constrained in both their spatial position (field or cluster) and their physical parameters (binarity, mass, age etc.).

\begin{acknowledgements}
  Based on observations at the European Southern Observatory in
  programme 074D.0011 and 171.D-0237. This work has made use of data from the
  European Space Agency (ESA) mission {\it Gaia}
  (\url{https://www.cosmos.esa.int/gaia}), processed by the {\it Gaia}
  Data Processing and Analysis Consortium (DPAC,
  \url{https://www.cosmos.esa.int/web/gaia/dpac/consortium}). Funding
  for the DPAC has been provided by national institutions, in
  particular the institutions participating in the {\it Gaia}
  Multilateral Agreement. CJE and DJL acknowledge the OWC.
  CJE thanks Linda Smith for obtaining the AAT spectra, and is
  grateful for the past hospitality of the Space Telescope Science
  Institute and travel funding from its Director's Discretionary
  Research Fund (DDRF).
\end{acknowledgements}

{ \tiny \bibliography{aa35415}}

\begin{appendix}
\onecolumn
\section{Data Tables}
{\tiny
\begin{center}
\begin{longtable}{lcccccll}
  \caption[]{Observational parameters of target stars.
    Cross-references to previous identifications are given in the
    final column as follows: \citet[][Sk]{san68},
    \citet[][AzV]{azz75,azz82}, \citet[][WB]{wal86a}, \citet[][NMC]{nie86},
    \citet[][MPG]{mas89}, \citet[][MA93]{mey93}, \citet[][KWB]{kel99}, and
    \citet[][2dFS]{eva04}. Additionally eclipsing binaries (EB) from the OGLE project \citep{uda98} have been identified. 
    Radial distances from \#1001 ($r_{\rm d}$,
    in arcmin) are given for each star (1$'$ corresponds to a physical
    distance of $\sim$17.5\,pc).
\label{346dat}}\\
\hline
ID & $\alpha$(2000) & $\delta$(2000) & r$_{\rm d}$ & $V$ & $B-V$ & Spectral Type & Comments \\
\hline 
\endfirsthead
\caption[]{\it{continued}} \\
\hline
Star & $\alpha$(2000) & $\delta$(2000) & r$_{\rm d}$ & $V$ & $B-V$ & Spectral Type & Comments \\
\hline 
\endhead
\hline 
\multicolumn{8}{r}{\it{continued on next page}} \\
\endfoot
\hline
\\
\multicolumn{8}{l}{$\dagger$: Further notes on individual stars:}\\
\multicolumn{8}{l}{{\it 1001:} \citet{wal78,wal86a}: O4 III(n)(f)}\\
\multicolumn{8}{l}{{\it 1002:} \citet{hum83}: B3 I; \citet{len97}: B1.5 Ia}\\
\multicolumn{8}{l}{{\it 1003:} \citet{hum83}: B3 I}\\
\multicolumn{8}{l}{{\it 1007:} \citet{mas95}: B1 III}\\
\multicolumn{8}{l}{{\it 1008:} \citet{wal86a}: O3 III (f$^\ast$); \citet{nie86}: O3 III(f); \citet{wal00}: ON3 III(f$^\ast$); \citet{wal02}: O2 III(f$^\ast$)}\\
\multicolumn{8}{l}{{\it 1010:} \citet{wal86a,nie86}: O5-6 V; \citet[][but noted as triple system]{mas09}: O5.5 III((f))}\\
\multicolumn{8}{l}{{\it 1011:} \citet{gar87}: O9 III; \citet{wal95}: O9 III}\\
\multicolumn{8}{l}{{\it 1012:} \citet{nie86}: O5-6: V; \citet{wal00}: O4-5 V((f))}\\
\multicolumn{8}{l}{{\it 1018:} \citet{gar87}: B0n }\\
\multicolumn{8}{l}{{\it 1019:} Classification adopted from \citet{wal00}, also classified as O7 If by \citet{gar87}.}\\
\multicolumn{8}{l}{{\it 1030:} \citet{hey10}: O5 V\,$+$\,OB}\\
\multicolumn{8}{l}{{\it 1052:} \citet{hey10}: O9.5 V}\\
\multicolumn{8}{l}{{\it 1060:} \citet{hey10}: O8 V}\\
\multicolumn{8}{l}{{\it 1079:} \citet{hey10}: B0-0.5 V}\\
\multicolumn{8}{l}{{\it 1103:} EIS photometry has a non-physical (negative) colour, adopted photometry is from \citet{mas89}.}\\
\multicolumn{8}{l}{{\it 1190:} MA93\#1169 is a blend of this and another star.}
\endlastfoot
\hline
1001 & 00 59 04.49 & $-$72 10 24.7  &  0.00  & 12.61 &  $-$0.28 & O4 If\,$+$\,O5-6  & Binary (SB2); WB\,1, NMC\,26, MPG\,435\,$-$\,{\it O5.5 If}; $\dagger$\\
1002 & 00 58 35.82 & $-$72 16 24.8  &  6.39  & 12.80 &  \pp0.01 & B1.5 Ia           & Sk 73, AzV 210; $\dagger$; small $v_{\rm r}$ shifts of $\sim$10\,\kms\\
1003 & 00 59 43.46 & $-$72 04 18.7  &  6.79  & 12.96 &  $-$0.14 & B3 Iab            & Sk 81, AzV 234; $\dagger$\\
1004 & 00 57 05.45 & $-$72 09 51.9  &  9.13  & 13.34 &  \pp0.01 & A2 Ib             &  \\
1005 & 00 57 04.09 & $-$72 08 57.0  &  9.33  & 13.38 &  $-$0.02 & A2 Ib             &  \\
1006 & 00 59 54.88 & $-$72 13 06.3  &  4.70  & 13.40 &  $-$0.09 & A2 Ib             &  MPG\,859, 2dFS\#5102\,$-$\,{\it B9 (Ib)}\\
1007 & 00 58 54.80 & $-$72 13 17.1  &  2.97  & 13.46 &  \pp0.03 & B1 Ia             & Binary (SB1); AzV 214, MPG\,293; $\dagger$\\
1008 & 00 59 00.75 & $-$72 10 28.1  &  0.29  & 13.52 &  $-$0.22 & O2 III(f$^\ast$)  & WB\,3, NMC\,29, MPG\,355\,$-$\,{\it O3 V((f$^\ast$))}; $\dagger$\\
\end{longtable}
\end{center}
}

\newpage
{
\begin{longtable}{llcccrrrrrrr}
\caption{Atmospheric parameters (\teff, \logg) and projected rotational velocities (\vsini), luminosities, current masses and ages for the NGC\,346 sample. The fourth column indicates the method used to estimate the effective temperature, viz.\ H08: taken from \citet{hun08a}; M06: taken from \citet{mok06}; A: adopted from effective temperature scales of \citet{tru07}; I: interpolated from these temperature scales; E: extrapolated from these temperature scales; M: luminosity class V assumed; Si: silicon ionization balance; He: \ion{He}{i} line at 4026\AA; He+: \ion{He}{ii} lines at 4541 and 4686\AA; L: literature. Two methods were used to estimate the \vsini: profile fitting (PF) and Fourier transform (FT). Targets marked with an asterisk appear to be overluminous for an assumed luminosity class V and their estimates should be treated with caution. SB1 in the final column indicates that significant radial velocity variables were detected while the double-lined spectroscopic binaries are designated as SB2 (and one tripled-lined system as SB3). Many of the Be-type stars from the FSMS were classified by \citep{eva06} as `(Be-Fe)' to indicate Fe\,{\scriptsize II} emission. This spectral qualifier has not been used further elsewhere (including for the Survey spectra here) but for completeness cf. the published types, these suffixes are included below.}
\label{t_346results}
\\ \hline
Star & Spectral Type & T$_{\rm eff}$&Method&$\log g$&\multicolumn{2}{c}{\vsini} & $r$ & $\log$ L&Mass & Age & Comment\\
     &               &               &       &          & PF & FT\\
     &               & (K)         &     & (dex)  &\multicolumn{2}{c}{(km\,s$^{-1}$)} & (\arcmin)   & (L$_{\sun}$)        &(M$_{\sun}$)& (Myr) \\
\hline
\\
\endfirsthead
\caption[]{\it{continued}} \\
\hline
Star & Spectral Type & T$_{\rm eff}$&Method&$\log g$&\multicolumn{2}{c}{\vsini} & $r$ &log L&Mass & Age & Comment\\
     &               &               &       &          & PF & FT\\
     &               & (K)         &     & (dex)  &\multicolumn{2}{c}{(km\,s$^{-1}$)} & (\arcmin)   & (L$_{\sun}$)           &(M$_{\sun}$)& (Myr)\\
\hline
\\
\endhead
\hline\multicolumn{12}{r}{{\it continued on next page}}\\\endfoot
\hline \endlastfoot
0001  &            O7 Iaf+   &  34100  &  M06  &  3.35  &  74   &  76   &  2.13  &  5.97  &  55.6  &  2.9  &  SB1   \\
0002  &            A2: Iab   &    --   &  --   &   --   &  --   &  --   &  5.55  &   --   &  --    &   --  &        \\
0003  &                G0:   &    --   &  --   &   --   &  --   &  --   &  3.36  &   --   &  --    &   --  &        \\
0004$^*$  &       Be (B1:)   &  27300  &  AM   &  3.00  &  266  &  258  &  7.22  &  5.16  &  22.4  &  6.6  &        \\
0005  &              A0 II   &    --   &  --   &   --   &  --   &  --   &  8.49  &   --   &  --    &   --  &        \\
0006  &                F2:   &    --   &  --   &   --   &  --   &  --   &  5.69  &   --   &  --    &   --  &        \\
0007  &         O4 V((f+))   &  42800  &  M06  &  3.95  &  120  &  113  &  0.56  &  5.54  &  38.4  &  2.4  &  SB1   \\
0008$^*$  &            B1e   &  27300  &  AM   &  3.05  &  299  &  295  &  5.74  &  4.96  &  17.6  &  7.8  &        \\
\end{longtable}
}
	
\twocolumn	\section{Comments on individual targets}\label{s_app}
	
	{\it \#1001 -- MPG\,435:} This is the visually brightest object in the
	main body of NGC\,346, first observed spectroscopically by
	\citet[][his `NGC\,346~No.\,1']{wal78} and classified as O4~III(n)(f).
	From consideration of its absolute magnitude, and from the offset of
	the stellar radial velocities compared to the nebular emission
	features, \citet{nie86} suggested it as a likely binary. Indeed,
	\citet{nie02} reported radial velocity variations in the system
	(classifying the spectra as O4f$+$O:), with a refined estimate of the
	period of 24.2\,d given by \citet{nie04}.  The FLAMES data, observed
	as part of the Field~A configuration, also reveal a massive companion.
	Note that this is probably not related to the multiple components
	reported by \citet{hey91}, which are at distances of greater than
	1\farcs5 (thus outside the Medusa fibre diameter).
	
\begin{figure}[h]
		\begin{center}
			\hspace{-0.85cm}
			\includegraphics[angle=270,scale=0.34]{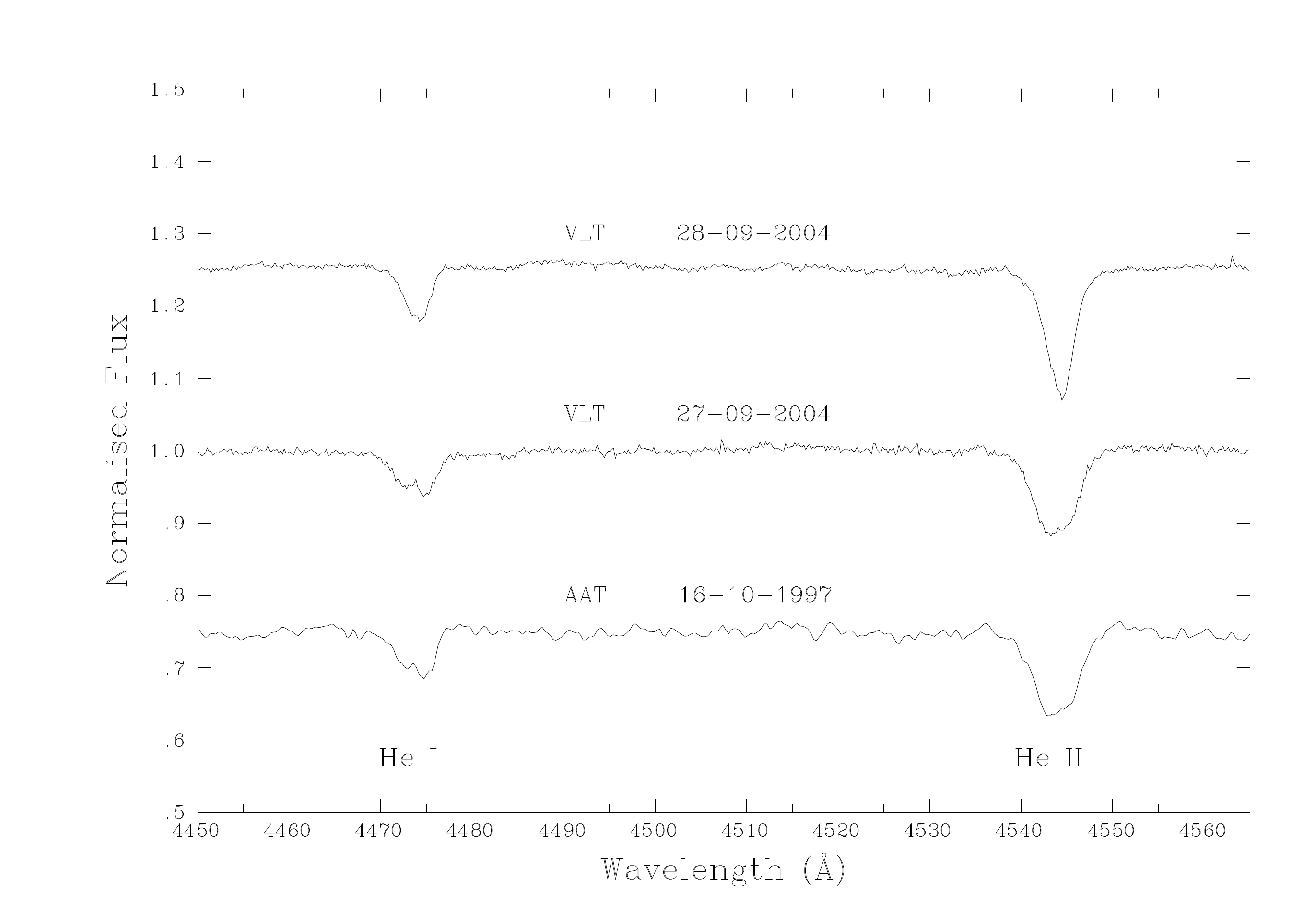}
			\caption{{\it \#1001 -- MPG\,435:} Multiple components are seen in the
				FLAMES spectra. Evidence for multiplicity is also seen in an
				unpublished AAT-UCLES spectrum (degraded and binned to the same
				resolution as the FLAMES data).}\label{1001}
		\end{center}
	\end{figure}

	The \lam\lam4450-4565 region of \#1001 is shown for our two LR02
	spectra in Fig.~\ref{1001}. Double components are clearly seen in the
	data from 2004 September 27, whereas the observation from the
	following night merely shows line asymmetry -- such a rapid change for
	a relatively long-period system is consistent with the eccentric orbit
	($e$\,$=$\,0.42$\,\pm$\,0.04) from \citet{nie04}.  For completeness we
	also include a previously unpublished spectrum of the system from the
	University College London Echelle Spectrograph (UCLES) on the Anglo
	Australian Telescope (AAT), contemporary to those presented by
	\citet{wal00} -- similar evidence of multiplicity is also seen.
	
	\ion{N}{iv} \lam4058 emission (requiring a spectral type of O4 or
	earlier) is seen in the LR02 spectra at a velocity consistent with the
	blueward component in Fig.~\ref{1001}.  The presence of \ion{He}{i}
	\lam4471 absorption, combined with weak \ion{He}{ii} \lam4686 emission
	(albeit complicated by absorption from the secondary) suggests a
	classification for the primary of O4~If. The \ion{He}{i} \lam4471
	absorption for the secondary is also weaker than that for \ion{He}{ii}
	\lam4542 (Fig.~\ref{1001}), suggesting a classification of O5-6.\\
	
	\vspace{-1.0cm}	
	
\begin{figure}[h]
		\begin{center}
			\hspace{-0.85cm}
			\includegraphics[angle=270,scale=0.34]{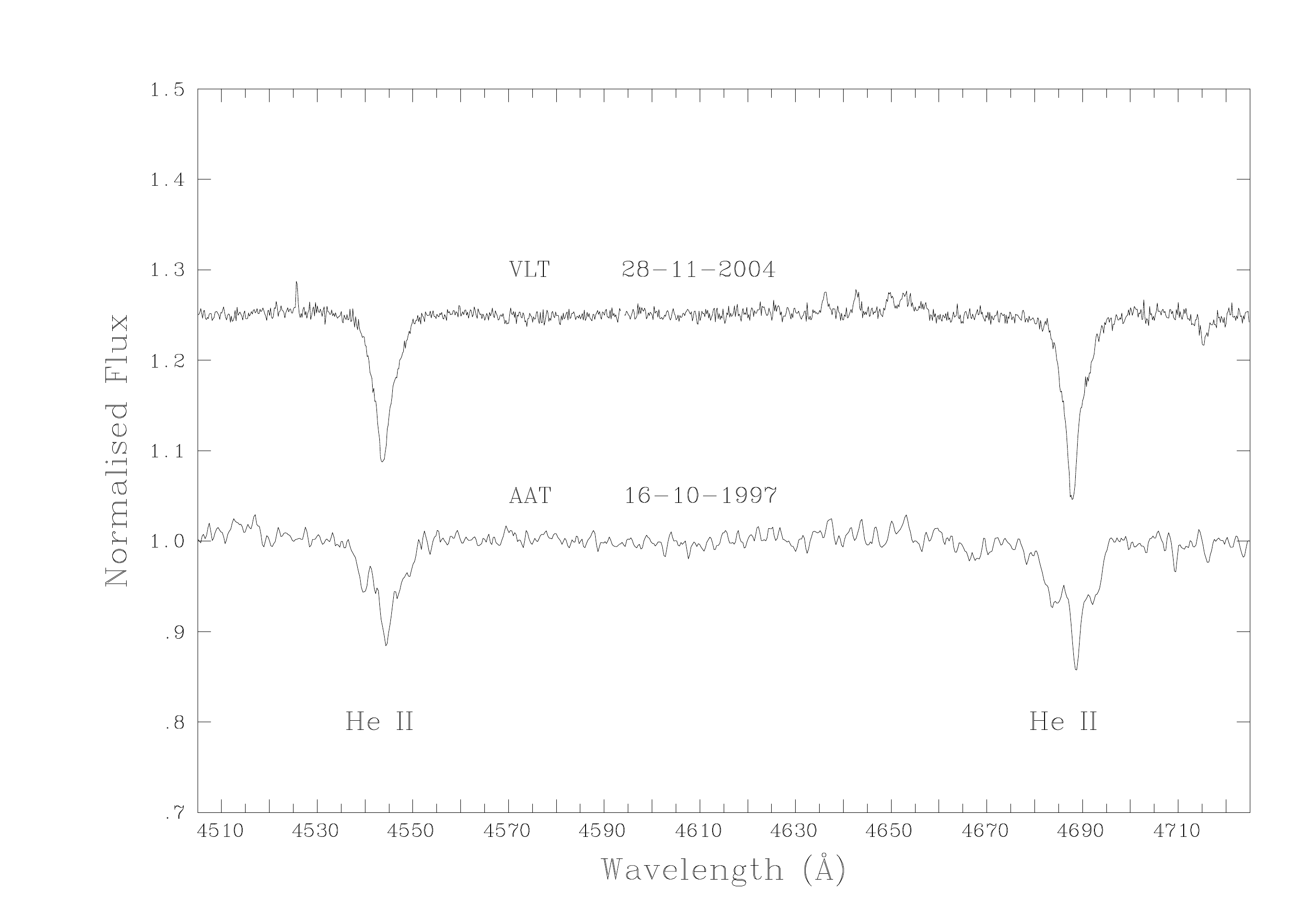}
			\caption{{\it \#1010 -- MPG\,342:} LR03 FLAMES spectrum in which
				\#1010 appears single, compared with a previously unpublished
				AAT-UCLES spectrum (degraded/binned to the FLAMES resolution), in
				which three components appear to be present.}\label{1010}
		\end{center}
	\end{figure}

	\noindent{\it \#1010 -- MPG\,342:} \citet{nie86} also suggested this star as a
	likely binary, with further monitoring by \citet{mas12} revealing a
	third component.  The FLAMES spectra, obtained in the Field~B
	configuration, are best described as O5-6~V((f)), but the multiplicity
	complicates precise classification and prompted us to re-inspect
	another unpublished AAT-UCLES spectrum \citep[again contemporary to
	those from][]{wal00} which also shows evidence of
	three components (see Fig.~\ref{1010}).\\
	
	\vspace{-1.0cm}	
	
	\noindent{\it \#1012 -- MPG 368:} As one of the brightest stars in the cluster,
	\#1012 has been observed by a number of authors. Classifications of
	O5-6:~V, O5.5~V((f$+$)), O4-5~V((f)) and O6~V were given by
	\citet{nie86}, \citet{mas89}, \citet{wal00}, and \citet{mas09},
	respectively. \citeauthor{mas09} reported a velocity difference
	(50\,\kms) between the He~\1 and \2 lines in their spectrum of the
	star and suggested it is a binary, with more recent high-resolution
	spectroscopy from \citet{bou13} finding evidence of two components in
	the He~\1 lines. The FLAMES spectra also reveal a velocity shift (of
	$\sim$50\,\kms) within the observations, providing further support of
	its binary nature (hence
	the past uncertainty in its classifications).\\
	
	\noindent{\it \#1024 -- MPG\,665, MA93\#1138, KWB346\#93:} This source was
	noted as an emission-line object by \citet[][their \#1138]{mey93} then
	by \citet[][\#93]{kel99} in their search for Be-type stars; on the
	basis of imaging polarimetry, \citet{wis07} argued that it is most
	likely a classical Be star. However, \#1024 was identified by
	\citet[][SMC\_SC8, \#160677]{uda98} as an eclipsing binary system from
	the second Optical Gravitational Lensing Experiment (OGLE) survey,
	with a period of 86.4\,d, updated to 86.2\,d by \citet{wyr04} and then
        86.25\,d by \citet{p13}.
	
\begin{figure}[h]
		\begin{center}
			\hspace{-0.85cm}
			\includegraphics[angle=270,scale=0.34]{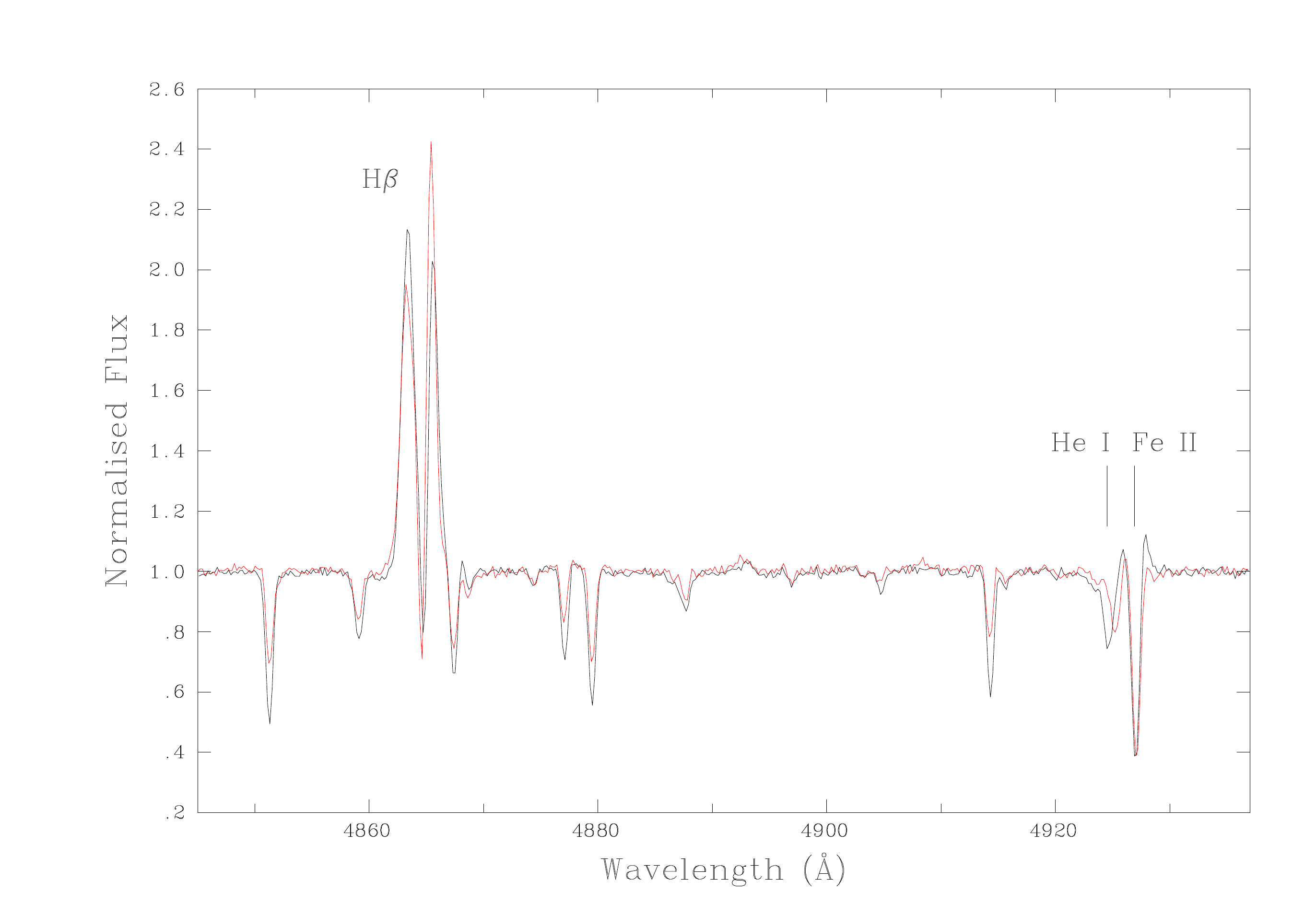}
			\caption{Combined LR03 observations for \#1024 from the two observational epochs.  The spectrum from the first epoch (black line) has been redshifted by 15\,\kms\/ such that the metallic spectra approximately agree -- note the offset in the \ion{He}{i} \lam4922 absorption, indicative of a hotter companion, and the shell-like emission of \ion{Fe}{ii} \lam4923.}\label{1024}
		\end{center}
	\end{figure}

	Strong twin-peaked emission is seen in H$\beta$, with weak \ion{He}{i}
	absorption combined with a rich metal-line absorption spectrum
	super-imposed (consistent with a late A spectral type). This initially
	suggested a Be-type object with a cooler disk. The time-sampling of
	our data is unfortunately relatively limited (\#1024 was observed in
	Field~A), but the LR03 observations were taken three nights apart and
	variations are seen in the H$\beta$ region (see Fig.~\ref{1024}). The
	first observation (2004 Oct 01) has been red-shifted by 15\,\kms\
	such that the metal absorption lines roughly match those in the second
	observation; note the residual offset in \ion{He}{i} \lam4922
	($\sim$40\,\kms) and shell-like emission in \ion{Fe}{ii} \lam4923.
	Similar behaviour is also seen in \ion{He}{i} \lam4713 and \lam5016
	but the velocity of the H$\beta$ emission appears consistent with the
	cooler spectrum. The blue:red ratio of the H$\beta$ emission changes,
	perhaps related to relative shifts in the line profile of the hotter
	companion rather than arising from the longer-term changes often seen
	and thought to arise from disk structure \citep{han95}. This system
	merits further study to ascertain its evolutionary status, but could
	be related to the peculiar A-type supergiants reported by \citet{men10}.\\
	
	\vspace{-1.0cm}	

	\noindent{\it \#1030 -- MPG\,655:} The spectrum of \#1030 (MPG\,655) has very
	narrow lines, similar to those seen in NGC\,346-028 from the FSMS
	({\it aka} MPG\,113), classified as OC6~Vz by \citet{wal00}.  The
	\ion{He}{i} lines suggest a slightly earlier type for \#1030, but
	there is also evidence of nebular infilling, so we classify its
	spectrum as OC5-6~Vz.  From lower-resolution spectroscopy,
	\citet{hey10} classified this star as O5~V\,$+$\,OB, with the
	suggestion of a secondary component arising from the strength of the
	\ion{He}{i} lines given an O5~V classfication (Dr.~N.~R. Walborn,
	private communication); from our (albeit limited) time coverage, we
	see no evidence for radial velocity shifts.  If the spectrum is
	convolved by a rotational broadening profile of, for example,
	200\,\kms, the \ion{He}{ii} \lam4686 line still remains stronger than
	the other helium lines, suggesting its classification as `Vz' spectrum
	is not due to either resolution or \vsini\/ effects.

\end{appendix}
\end{document}